\documentclass[useAMS,usenatbib]{mn2e}
\usepackage{graphicx, rotating}
\usepackage{aas_macros}  
\usepackage{graphicx}
\usepackage{amssymb}
\usepackage{rotating}
\usepackage{multirow}
\usepackage[breaklinks,colorlinks=true,citecolor=black,linkcolor=black,urlcolor=blue,bookmarks=true]{hyperref}
\usepackage{breakurl}  

\newcommand{\csq}{$\chi^2$}
\newcommand{\kms}{km\,s$^{-1}$}
\newcommand{\Reff}{R$_e$}

\begin{document}
\title[Star-formation histories in dEs]{Formation and Evolution of
  Dwarf Elliptical Galaxies - II. Spatially resolved star-formation histories
} \author[Mina Koleva et
  al.]{Mina Koleva$^{1,2}$\thanks{E-mail:
    mina.koleva@obs.univ-lyon1.fr}, Sven De Rijcke$^{3}$,
  Philippe Prugniel$^{1}$, Werner~W. Zeilinger$^{4}$,\newauthor
  Dolf Michielsen$^{3}$,
\\ $^{1}$Universit\'e Lyon~1,
  Villeurbanne, F-69622, France; CRAL, Observatoire de Lyon, St. Genis
  Laval, F-69561, France ; CNRS, UMR 5574 \\ 
  $^{2}$Department of
  Astronomy, St. Kliment Ohridski University of Sofia, 5 James
  Bourchier Blvd., BG-1164 Sofia, Bulgaria\\ 
  $^{3}$Sterrenkundig
  Observatorium, Ghent University, Krijgslaan 281, S9, B-9000 Ghent,
  Belgium\\
  $^{4}$Institut
  f\"{u}r Astronomie, Universit\"{a}t Wien, T\"{u}rkenschanzstrasse
  17, A-1180 Wien, Austria\\ 
  }
\date{Accepted 2009 03 23.  Received 2009 03 16; in original form 2008 11 19}

\pagerange{\pageref{firstpage}--\pageref{lastpage}} \pubyear{2009}

\maketitle 

\label{firstpage}

\begin{abstract}
We present optical VLT spectroscopy of 16 dwarf
elliptical galaxies (or dEs) comparable in mass to NGC~205, and belonging to
the Fornax cluster and to nearby groups of galaxies. 
Using full-spectrum fitting, we derive radial profiles of 
the SSP-equivalent ages and metallicities. We
make a detailed analysis with \textsc{ulyss} and \textsc{steckmap} of the star-formation history 
in the core of the galaxies and in an aperture of one
effective radius. We resolved the history into 1 to 4 epochs.
The statistical significance of these reconstructions
were carefuly tested; the two programs give remarkably consistent results.

The old stellar population of the dEs, which dominates their mass, is
likely co-eval with that of massive ellipticals or bulges, but the
star formation efficiency is lower.  Important intermediate age (1-5
Gyr) populations, and frequently tails of star formation until recent
times are detected. These histories are reminiscent of their lower
mass dSph counterparts of the Local Group.

Most galaxies (10/16) show significant metallicity gradients, with metallicity
declining by 0.5~dex over one half-light radius on average. These gradients
are already present in the old population.
The flattened (or discy), rotating objects (6/16) have flat metallicity
profiles. This may be consistent with a distinct origin for these galaxies
or it may be due to their geometry.  
The central SSP-equivalent age varies between 1 and 6 Gyr, with the
age slowly increasing with radius in the vast majority of objects. The
group and cluster galaxies have similar radial gradients and
star-formation histories.

The strong and old metallicity gradients place important constraints
on the possible formation scenarios of dEs. Numerical simulations of
the formation of spherical low-mass galaxies reproduce these
gradients, but they require a longer time for them to build up. A
gentle depletion of the gas, by ram-pressure stripping or starvation,
could drive the gas-rich, star-forming progenitors to the present dEs.
 
\end{abstract}

\begin{keywords}
  galaxies: dwarf - galaxies : formation and evolution - galaxies : stellar
  populations
\end{keywords}

\section{Introduction}

Diffuse ellipticals, or dwarf elliptical galaxies (dEs), are small,
low-luminosity galaxies ($M_\textit{B} \gtrsim -18$~mag)
\citep{fb94}. They are among the most numerous galaxy species in the
universe. Being found typically not more than a few hundred
kiloparsecs away from a massive galaxy or in groups and clusters of
galaxies, they show a strong predilection for high-density
environments. Their diffuse, approximately exponentially declining
surface-brightness profiles set them apart from the compact
ellipticals (cEs) which, while occupying the same luminosity range as
the dEs, have much higher central surface brightnesses and a de
Vaucouleurs-like surface-brightness profile
(\citealp{kormendy85,np87,kormendy08}; but see \citealp{graham03}).

The star formation and metal enrichment history are likely not
regular, as star formation may be both triggered and quenched by
environmental effects \citep{m01,h07,vandenbosch08}.  Moreover, dEs
may not be a homogeneous population. There is tempting evidence that
at least part of the dE population evolved from more late-type,
gas-rich progenitors. This includes the occurrence of dEs with a
rotation velocity that is compatible with them being isotropic oblate
rotators \citep{ddzh01,ps023,vz04} and dEs with kinematically
decoupled cores and/or embedded stellar discs and bars
\citep{j00,b02,g03,d03,d04,cpsa07,lt07}. High-speed gravitational
interactions with giant galaxies within the dense group or cluster
environment may be sufficient to induce the morphological
transformation of a discy dwarf irregular galaxy into a much rounder
dwarf spheroidal (dSph) or dwarf elliptical galaxy
\citep{m98,m01,m05}. Some dEs still contain an interstellar medium
\citep{yl97,dzdh03,mdzpdr04,Buyle05,bouchard07} and host ongoing star formation
\citep{dzdh03,lt06,michielsen08}

Detailed information about the star formation histories of a
representative sample of dEs is instrumental in elucidating their
origin and evolution. In the first paper of this series (\citealp{d05},
Paper I), we compared simulations of the evolution of dEs with
observed scaling relations, such as the Faber-Jackson relation and the
Fundamental Plane. We found good agreement with models based on the
premise that dEs are primordial objects whose dark-matter halo merging
time-scale is shorter than the star-formation time-scale and in which
supernova explosions play an important role in regulating star
formation and moving gas around. Part of the data used in Paper I was
collected in the course of an ESO Large Programme on the internal
kinematics of dEs (program ID 165.N-0115). In the present paper, we
use deep long-slit spectra to probe the composition of the stellar
populations and the star-formation histories of the same sample of dEs
as in Paper I. This sample contains dEs with kinematically decoupled
cores and embedded discs and spiral structures whose stellar
populations can be compared with those of prima facie primordial dEs.

In section \ref{obsred}, we describe the sample of 16 objects, the
observations, and the data reduction. In section \ref{radpop}, we
present the radial profiles of SSP-equivalent ages and
metallicities. Section \ref{sec:sfh} contains a detailed analysis of
the stellar populations in the nuclei and main bodies of the targeted
galaxies and a reconstruction of the star-formation histories in these
two zones. We finalize with a discussion of the results in section
\ref{disc} and summarize our conclusions in section \ref{conc}.

\section{Observations and data reduction} \label{obsred}

\subsection{The sample}

\begin{table*}
\centering
\begin{minipage}{180mm}
\caption{Structural properties of the observed dEs (taken from \citet{d05}).}
\label{table:1}
\begin{tabular}{|l|cc|c|c|c|c|c|c|l|l} \hline
 name    & $\alpha$ (J2000) & $\delta$ (J2000) & $m_B$  &  $\epsilon$\footnote{$\epsilon = 1-b/a$, with $b/a$ the  minor to major axis ratio of the isophotes}
         &  $\sigma$     & $R_{\rm e}$  & $\mu_{{\rm e},B}$ & S\'ersic $n$  & type & comment \\ 
         & (hms) & (dms) & (mag) &  & ({\kms}) & (arcsec) & (mag arcsec$^{-2}$) & & & \\\cline{1-11}
  FCC136 & 03 34 29.5       &  -35 32 47       & 14.81  &   0.21       & 64.3$\pm$3.8  &  14.2       &  22.57& 1.71 & dE2   &                   \\
  FCC266 & 03 41 41.4       &  -35 10 12       & 15.90  &   0.11       & 42.4$\pm$3.4  &   7.1	     &  22.15& 1.08  & dE0,N &                  \\
  FCC150 & 03 35 24.1       &  -36 21 50       & 15.70  &   0.19       & 63.8$\pm$3.9  &   5.7       &  21.48& 1.65 & dE4,N &          \\
  FCC245 & 03 40 33.9       &  -35 01 23       & 16.00  &   0.11       & 39.5$\pm$4.2  &  11.4       &  23.28& 1.35  & dE0,N &                  \\
  NGC5898\_DW1 & 15 18 13.0 &  -24 11 47       & 15.66  &   0.34       & 43.5$\pm$3.0  &   8.7       &  22.35& 1.53  & dE3   &          \\
  FS029  & 13 13 56.2       &  -16 16 24       & 15.70  &   0.54       & 59.6$\pm$3.6  &   8.9       &  22.44& 1.79  & dE5   &                   \\
  FS373  & 10 37 22.9       &  -35 21 37       & 15.60  &   0.23       & 73.1$\pm$3.1  &   7.9       &  22.03& 2.71 & dE3   &KDC                \\
  FS076  & 13 15 05.9       &  -16 20 51       & 16.10  &   0.07       & 56.8$\pm$3.8  &   4.4       &  21.41& 2.02 & dE1   &KDC                \\
  FCC288 & 03 43 22.6       &  -33 56 25       & 15.10  &   0.72       & 48.5$\pm$3.3  &   9.5       &  21.99& 1.11  & dS0   &spiral             \\
  FS131  & 13 16 49.0       &  -16 19 42       & 15.30  &   0.54       & 87.0$\pm$3.2  &   8.1       &  21.83& 2.35  & dE5,N &peanut          \\
  FS075\footnote{not observed in ESO Large Programme 165.N-0115; the velocity dispersion is a preliminary value obtained with the current data, its relatively large error bar is due to the uncertainty on the instrumental line-spread function.}%
         & 13 15 04.1       &  -16 23 40       & 16.87  &   0.10       & 49$\pm$6          &   6.8       &  23.03& 1.74  & dE1,N &                  \\
  FCC207 & 03 38 19.3       &  -35 07 45       & 16.19  &   0.15       & 60.9$\pm$6.6  &   8.4       &  22.81& 1.32  & dE2,N &H$\alpha$ emission\\
  NGC5898\_DW2 & 15 18 44.7 &  -24 10 51       & 16.10  &   0.57       &  44.2$\pm$3.4 &   5.9       &  21.95& 1.26  & dS0,N   &emission      \\ 
  FCC204 & 03 38 13.6       &  -33 07 38       & 14.76  &   0.61       & 67.2$\pm$4.4  &  11.5       &  22.06& 1.29  & dS0   &spiral             \\
  FCC043 & 03 26 02.2       &  -32 53 40       & 13.91  &   0.26       & 56.4$\pm$3.7  &  16.9       &  22.05& 2.17  & dE3   &                   \\
  FCC046 & 03 26 25.0       &  -37 07 41       & 15.99  &   0.36       & 61.4$\pm$5.0  &   6.7       &  22.12& 1.24  & dE4,N &H{\sc i}, H$\alpha$ emission\\
\cline{1-11}
\end{tabular}
\end{minipage}
\end{table*}

The sample consists of 16 dEs, 9 in the Fornax cluster, 4 in the NGC~5044
group, two in the NGC~5898 group and one in the Antlia (NGC~3258) group.
Structural properties of the objects are listed in Table~\ref{table:1}. 
They are essentially comparable to NGC~205, the most luminous being
about 6 times brighter than this prototype.
The galaxies are ordered according to the importance of their old population,
as determined later in this study (Sect.~\ref{subsec:sfh}), 
starting with those most dominated by the old population (i. e. the shortest 
formation time-scale).

We have collected photometric data of dSphs/dEs from the             
galaxy groups and clusters represented in our sample. Analogous to 
\cite{gyf03}, we quantify the galaxy densities of these environments 
as the projected density of galaxies brighter than M$_B=-19$~mag              
within the radius $d_5$ that contains five such galaxies, or:
\begin{equation}
\Sigma_5 = \frac{5}{\pi d_5^2}\,{\rm galaxies}\,{\rm Mpc}^{-2}. \label{form}
\end{equation}
The $\log(\Sigma_5)$-values and other characteristics of the different 
environments are listed in Table \ref{tab:dens}.

\begin{table}
 \centering
 \begin{minipage}{\columnwidth}
  \caption{Characteristics of the galaxy groups  from which
    the sample was composed.
    Columns 2 \& 3: Mean radial velocity relative to the CMB and
    velocity dispersion listed in HyperLeda, http://leda.univ-lyon1.fr, 
    \citep{HYPERLEDA} calculated as in \citet{prugniel99}.
    Columns 4 \& 5: Distance modulus, assuming H$_0$ = 72 \kms/Mpc, and
    radial scale for 1 arcsec at the corresponding distance.
    Column 6: Central bright-galaxy density, measured by $\log(\Sigma_5)$
    (see eq. (\ref{form})). \label{tab:dens}}
  \begin{tabular}{@{}lccccc@{}}
  \hline
 group/cluster & cz   & $\sigma$&mod& r  & $\log(\Sigma_5)$  \\
               & \kms & \kms &    & pc &                   \\
 \hline
NGC5898 group   & 2390 & 145 & 32.60 & 160 & -0.7 \\
Fornax cluster  & 1540 & 285 & 31.65 & 104 &  0.5 \\
NGC5044 group   & 2700 & 210 & 32.87 & 182 &  0.7 \\
NGC3258 group   & 2860 & 280 & 33.00 & 193 &  1.4 \\
 \hline
\end{tabular}
\end{minipage}
\end{table}

The NGC5989 group constitutes the sparsest environment covered by the
dataset (it consists of two bright ellipticals, NGC5903 and NGC5898,
and a few tens of much fainter galaxies; \citealt{gcf92} list only
three group members brighter than M$_{\textit{B}} \approx
-18$~mag). The Fornax cluster and the NGC5044 group have comparable
bright galaxy densities and can be considered to be similar
environments, significantly denser than the NGC5898 group. The Antlia
cluster has a densely populated core, reflected by its high
$\log(\Sigma_5)$ value.

The spectroscopic observations were carried out during two ESO-VLT
service mode runs. The group dEs were observed in April -- May 2005
using FORS2 and the Fornax dEs were observed in December 2005 --
January 2006 using FORS1.  The setup of the observations is summarized
in Table~\ref{table:2}. Exposure times are listed in
Table~\ref{table:3}.  During the observations, the seeing conditions
varied from FWHM = 0.5 to 0.8 arcsec.  We also observed three velocity
and metallicity standards for each run; two of which belong to the
Lick/IDS and the MILES stellar libraries \citep{so6}. The data
reduction, described in detail below, was performed using the
ESO-MIDAS data reduction software package.

\begin{table}
\caption{Setup of observations }
\begin{tabular}{lcc}
\hline
                                         & FORS 1       & FORS 2        \\
\hline
VLT unit                                 & Kueyen (UT2) &  Antu (UT1)   \\
CCD                                      & Tektronix    & 2 MIT mosaic  \\   
\# of pixels                             & 2k$\times$2k & 2k$\times$4k  \\      
                                         &              &(2$\times$2 binned)   \\     
pixel size (${\mu}$m$^2$)                & 24$\times$24 & 30$\times$30  \\             
image scale (arcsec pix$^{-1}$ )            & 0.20         & 0.25         \\     
readout noise (e$^-$ pix$^{-1}$ )         & 5.6          & 3.15         \\      
gain (ADU (e$^-$ )$^{-1}$ )               & 0.71         & 1.43         \\         
FORS grism                               & GRIS\_600B+22& GRIS\_1200g+96\\         
slit width (arcsec)                        & 0.5          & 1.0           \\         
spectral range (\AA)                     & 3300 -- 6200 & 4355 -- 5640  \\      
FWHM $\Delta\lambda$ (\AA)               & 2.6          & 3.0           \\     
$\sigma_{\rm instr}$ ({{\kms}} @ 5200 \AA) & 64           & 74 \\             
\hline
\end{tabular}
\label{table:2}
\end{table}

\subsection{FORS2 observations}

The calibration data include bias frames, dome flat-field frames, and
wavelength calibration frames which were taken during the day before
and after the observations. During twilight of each night,
spectrophotometric standard stars were observed, using a 5~arcsec
slit. For three nights we did not have calibration data, in which case
we took the frames from the following night. The bias frames for each
night (typically 10) were averaged and then median filtered over a box
of $5\times 5$~pixels. The dome flat frames (typically 5) were bias
subtracted, normalized and averaged.  The resulting frame was median
filtered over 40 pixels in the wavelength direction in order to yield
the variation of the intensity of the dome lamp with wavelength. The
small-scale flat-field was computed by dividing the averaged by the
median filtered frame. We also took twilight flats which were bias
subtracted, flat-fielded, average combined and averaged along the
wavelength direction to measure the variation of the illumination
along the slit (spatial direction).  Because of the very high S/N
there was no need to approximate this function with a
polynomial. The small-scale flat-field was multiplied by the
normalized slit transfer function to yield a master flat frame.

      The wavelength calibration was done on He-Hg-Cd arc frames observed with
the same setup.  Fitting a two-dimensional dispersion
relation using two-dimensional 4$^{th}$-order polynomials yielded an overall
accuracy of 0.06~\AA{} r.m.s. or $\sim$ 3.6~{\kms} at 5000~\AA.  The
wavelength calibrated frames were all rebinned to a common 1.5~\AA{}
pix$^{-1}$ wavelength scale.

      The FORS2 data show quite strong distortions of more than 3 pixels 
over the whole wavelength range.
In order to correct it, we determined the variation of the
spatial centre of the bluest standard star, and fitted a 3$^{rd}$ order
polynomial along the wavelength range. We chose the bluest standard star
because the quantum efficiency of the CCD in the blue part is not very high and
the centre of the objects are very difficult to determine except in this
star. Applying this correction to the other spectra, we noted a significant
residual linear distortion. This was fitted in the red part of the spectrum
where the centre is well defined and then added to the original 3$^{rd}$ order
polynomial to make one resampling correcting the distortion to better than
0.02~arcsec r.m.s.

     Each science frame was bias-subtracted and flat-fielded.  Then we removed
the cosmics by first applying the MIDAS FILT/COS command and subsequently
inspecting all the frames by eye to remove low-intensity cosmics missed by the
routine. Then we corrected for the slit distortion as described above. After
that the frames were wavelength calibrated and rectified. Sky subtraction was
performed by fitting the regions of the spectrum empty of target light
contributions with a linear relation. Each frame was corrected for extinction
using the extinction coefficients provided by ESO and the mean of the airmass
at the start and end of each exposure.  The spectra were flux-calibrated using
the spectrophotometric standard stars and spectral images of the same object
were co-added.

     The error spectra were calculated by transporting the photon and 
read-out noises through all the steps of the data reduction.

\subsection{FORS1 observations}

The reduction of the FORS1 data is identical to that of the FORS2
observations, except for the following steps: For the wavelength calibration we
used 12 lines. We fitted a
two-dimensional dispersion relation using polynomials of order 3 and 5 in
wavelength and spatial scale respectively, to an accuracy of 0.03~\AA{} rms or
$\sim$ 2~{\kms} at 5000~\AA. The frames were rebinned
to 0.6~\AA{} pixel$^{-1}$ .  The FORS1 CCD shows negligible distortion ($<$
0.05~arcsec), which was not corrected.

\begin{table}
\centering
\begin{minipage}{\columnwidth}
\caption{Observation log}
\label{table:3}
\begin{tabular}{lcrr} \hline 
\multicolumn{1}{l}{galaxy name} & \multicolumn{1}{c}{FORS} &
      \multicolumn{1}{c}{t$_{\rm obs}$} & \multicolumn{1}{c}{P.A.}\\
\multicolumn{1}{c}{ }  & \multicolumn{1}{c}{(1/2)}&
      \multicolumn{1}{c}{(s)} & \multicolumn{1}{c}{slit ($^\circ$)}  \\
\hline              
FCC 043    &   1 &  4$\times$960  &  80 \\
FCC 046    &   1 &  3$\times$1900 &  80 \\
FCC 136    &   1 &  4$\times$960  & 172 \\
FCC 150    &   1 &  3$\times$1900 &  10 \\
FCC 204    &   1 &  3$\times$1900 &  22 \\
FCC 207\footnote{\small 1 exposure excluded (clouds)}              &  1 & 4$\times$2280 & 114 \\
FCC 245\footnote{\small 1 exposure excluded (seeing $> 2$~arcsec FWHM)} &  1 &14$\times$1300 &  10\\
FCC 266    &   1 &  4$\times$2280 & 10 \\
FCC 288\footnote{\small 1 exposure excluded (clouds)}              &  1 & 4$\times$2280 &   4\\
FS 29      &   2 &  6$\times$ 1300&  98\\
FS 75      &   2 &  4$\times$1300 &   0\\
FS 76      &   2 &  4$\times$1300 &  36\\
FS 131     &   2 &  4$\times$1300 & 131\\
FS 373     &   2 &  4$\times$1300 &  97\\
NGC 5898 DW1 & 2 & 10$\times$1300 & 174\\
NGC 5898 DW2 & 2 & 14$\times$1300 &  12\\
\hline
\end{tabular}
\end{minipage}
\end{table}

\subsection{Analysis of the individual exposures}

Before co-adding the different exposures we analysed their central
1~arcsec extraction in order to check their internal consistency. In
order to minimize the change of the line-spread function (LSF, the
analogous of the PSF in the wavelength direction), we fitted the
line-of-sight velocity distribution (LOSVD), the age and the metallicity 
in a short wavelength range (4310 -- 4400\AA{} in rest-frame). 
We identified some exposures
producing slightly discrepant results, mostly for the LOSVD but also
for the population parameters.  Checking
the ambient conditions from the ESO meteo monitoring, we found that
these exposures were taken in worse seeing conditions than requested.

Only for 3 galaxies did the derived kinematics and stellar population
parameters differ significantly from frame to frame. In one case -
FS29 - the seeing changed from 0.6 arcsec to almost 1.8 arcsec
FWHM over the series of exposures. We excluded therefore the last
exposure of the series, in order to preserve the high resolution in
the co-added spectrum.  For the other 10 galaxies the seeing
variations were, although measurable, not significant.

We also notice that many exposures (especially for FORS2 observations)
were taken with a seeing smaller than the slit width. As shown by the
different instrumental broadening measured on the twilight spectra
(uniform lightening of the slit) and on the calibration stars, this
has a significant effect on the measurement of the LOSVD. As we will
see later, our analysis methods fits in the same time the LOSVD and
the parameters of the population. 
In the present case, the velocity and velocity dispersion are likely biased
and we will not discuss them. However, in order to suppress the
metallicity -- velocity dispersion degeneracy \citep{koleva08b}, the
LOSVD will be free parameters of our analysis.

\subsection{Extraction of 1D spectra}

For each object, we extract two 1D spectra integrated within simulated elliptical 
apertures with major axes
equal to 0.5~arcsec and 1 effective radius, \Reff.

Assuming the slit width to be
negligible, a spectrum at a distance $r_i$ along the slit is supposed
to be representative of the semi-ellipse with major axis $r_i$ and
with ellipticity $\epsilon$ (for which we take the galaxy's mean
ellipticity). If each pixel measures $\Delta r$ arcseconds on the sky,
a spectrum at a distance $r$ from the galaxy centre is then weighted
with a factor $2 \pi r (1-\epsilon) \Delta r$ in the extracted
spectrum, $\cal F$. A radial bin at radius $r_i$ starts at $r_i -
0.5{\Delta}r$ and ends at $r_i + 0.5{\Delta}r$. We then find for the
extracted spectrum that
\begin{equation}
{\cal F}_\lambda = 2 \pi \Delta r \sum_i F_{i,\lambda} r_i, 
\end{equation}
with $F_{i,\lambda}$ the observed spectrum at a distance $r_i$ from
the centre, and in the noise on the extracted spectrum, $\cal N$,
\begin{equation}
{\cal N}_\lambda = \sqrt{ \sum_i (2\pi \Delta r)^2 r_i^2 N_{i,\lambda}^2},
\end{equation}
with $N_{i,\lambda}$ the noise on the observed spectrum at a distance
$r_i$ from the centre. 

The central extraction is approximately the seeing disc and
corresponds to a diameter between 104 and 182 pc for the various
objects of the sample (see Table~\ref{tab:dens}). This extraction is
likely essentially sensitive to the population in the core, while the
external extraction, within 1 \Reff{} and excluding the inner 1~arcsec
region, can be considered as representative of the galaxy's main body.

\section{Radial profiles of the population parameters} \label{radpop}

In this section, after describing the method used to determine them,
we present the radial profiles of the age and metallicity of the galaxies.
Then we measure these characteristics in the central regions, after
correction of the light contamination projected along the line of sight.

\begin{figure*}
\includegraphics[height=0.33\textheight]{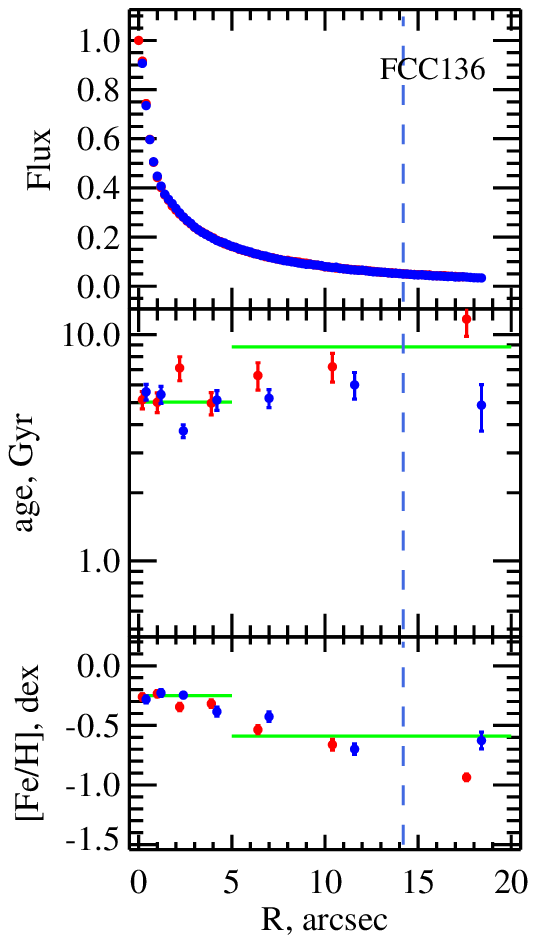}
\includegraphics[height=0.33\textheight]{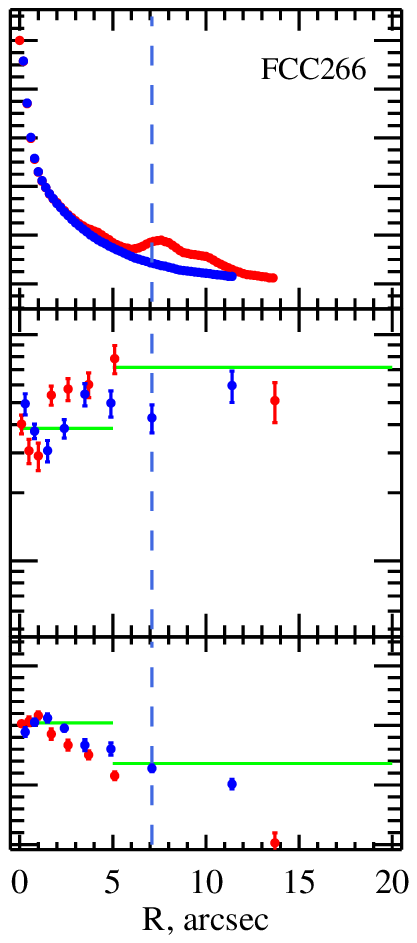}
\includegraphics[height=0.33\textheight]{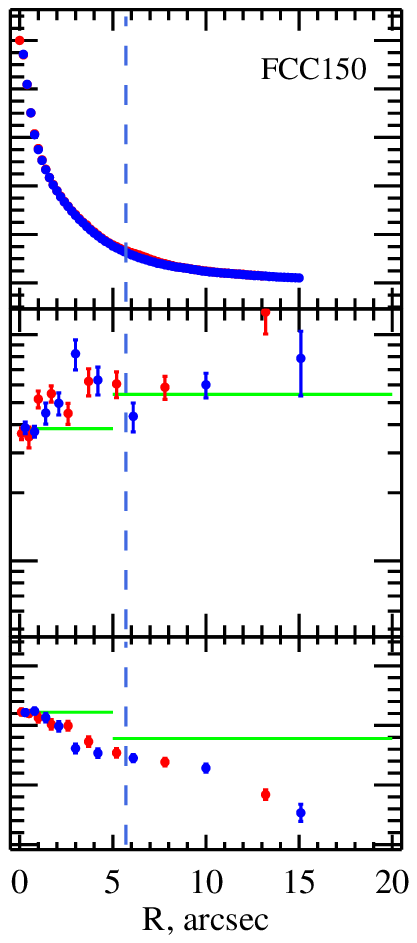}
\includegraphics[height=0.33\textheight]{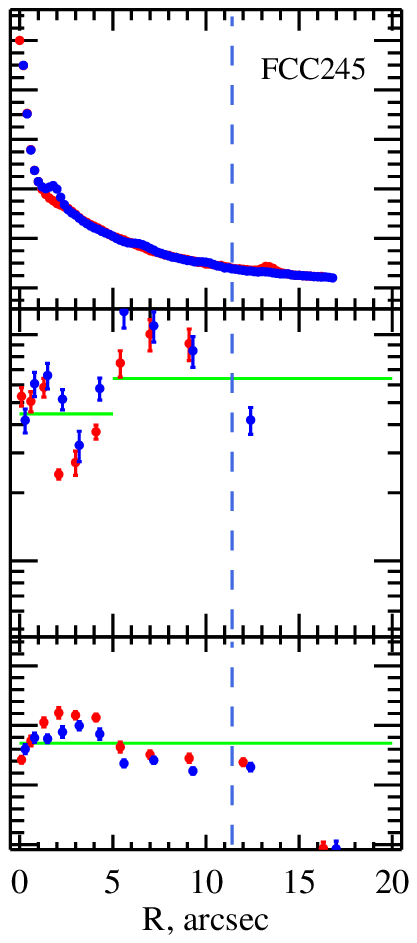}
\includegraphics[height=0.33\textheight]{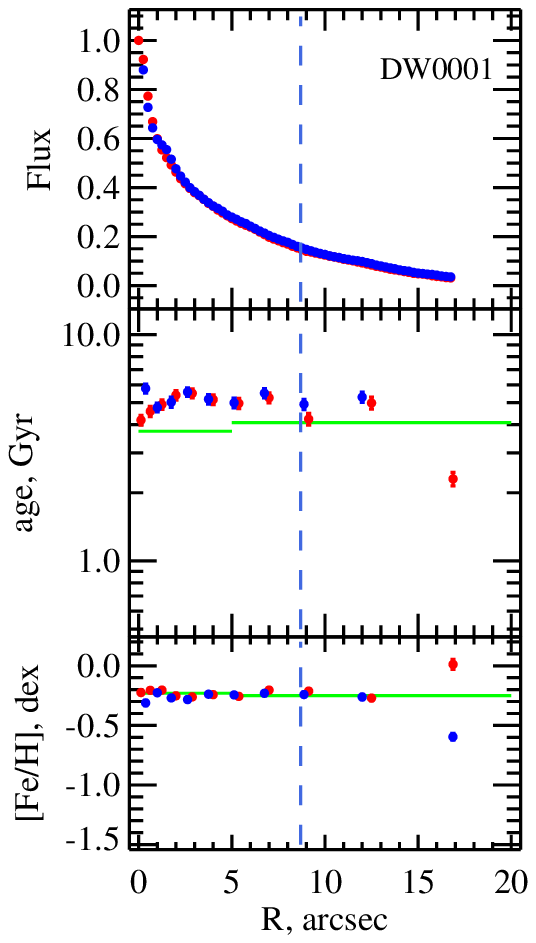}
\includegraphics[height=0.33\textheight]{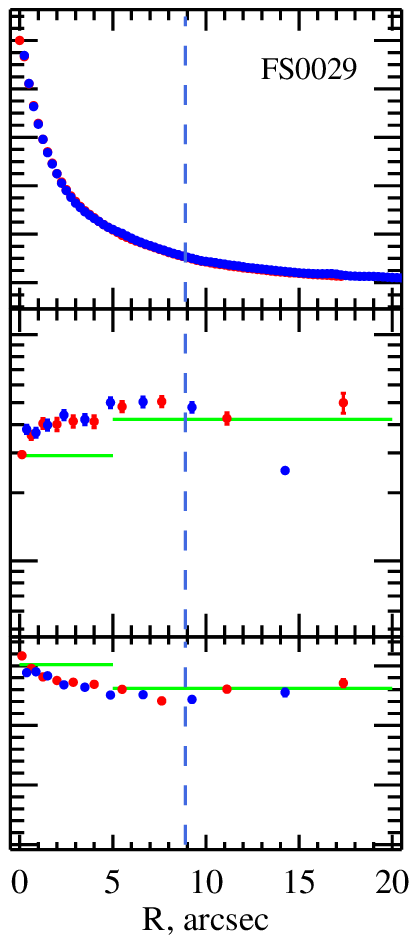}
\includegraphics[height=0.33\textheight]{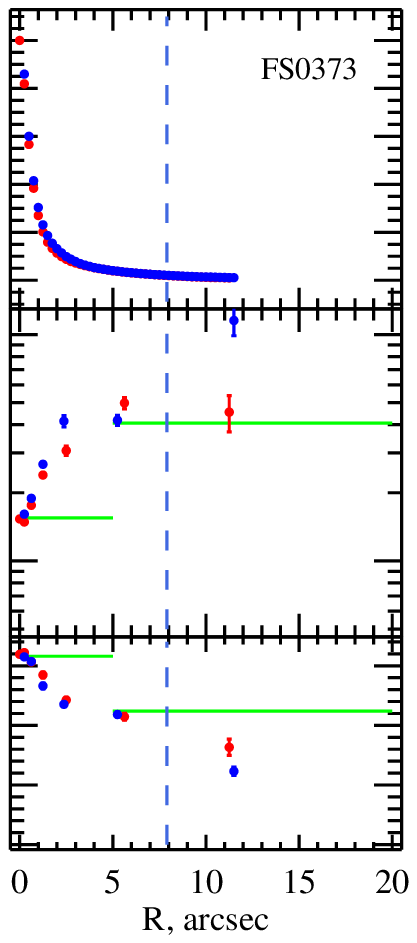}
\includegraphics[height=0.33\textheight]{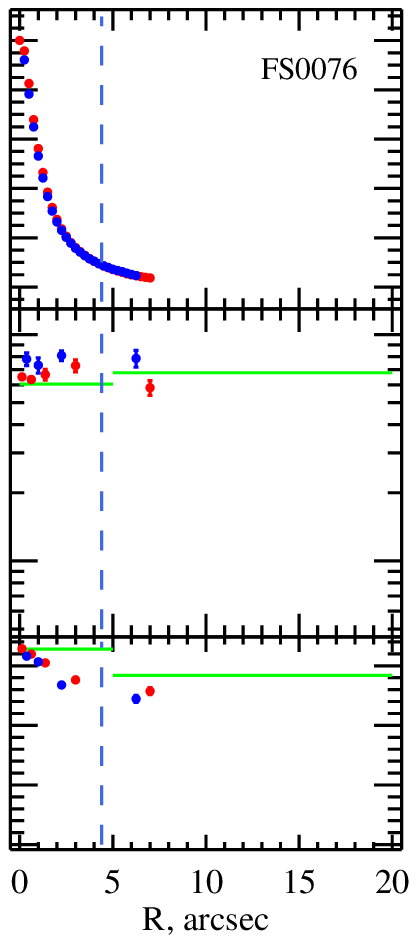}
\caption{ Radial profiles of the dEs with the shortest star-formation
  time-scale (see paragraph \ref{subsec:sfh}). In the outer regions we
  binned the spectra to have SNR $>$ 20. One half-light radius, \Reff, is
  indicated with a vertical blue dashed line. The SSP-equivalent
  metallicities and ages within the 1~arcsec (uncorrected
  for line-of-sight contamination) and 1~\Reff{} extractions
  (Sect. \ref{sec:sfh}) are shown with green lines 
  (for readability these lines are arbitrarily drawn from 0 to 5 arcsec
  and from 5 to 20 arcsec). The profiles are
  folded around the luminosity peak, the red points are for the
  'right' side of the centre (North or East), and the blue for the
  'left'.  }
\hskip 0cm
\label{fig:radprof1}
\end{figure*}

\begin{figure*}
\includegraphics[height=0.33\textheight]{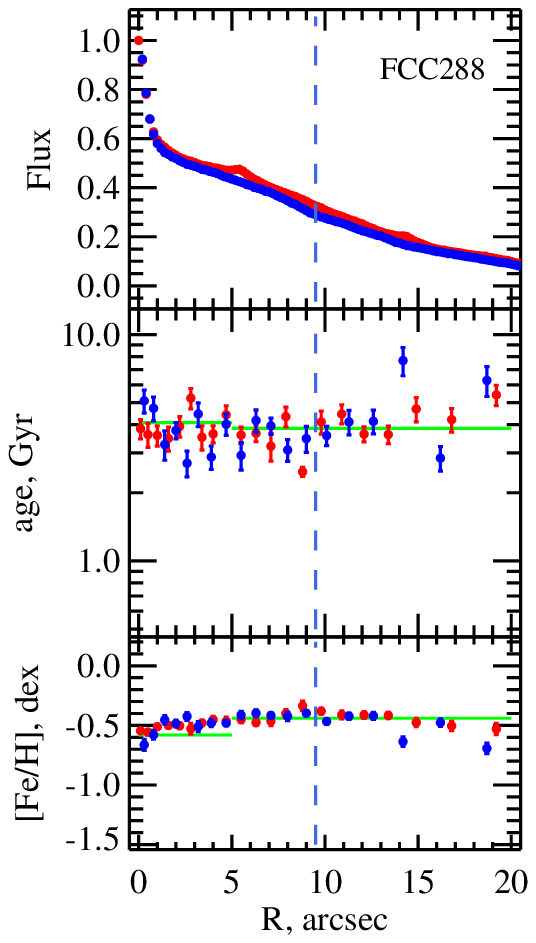}
\includegraphics[height=0.33\textheight]{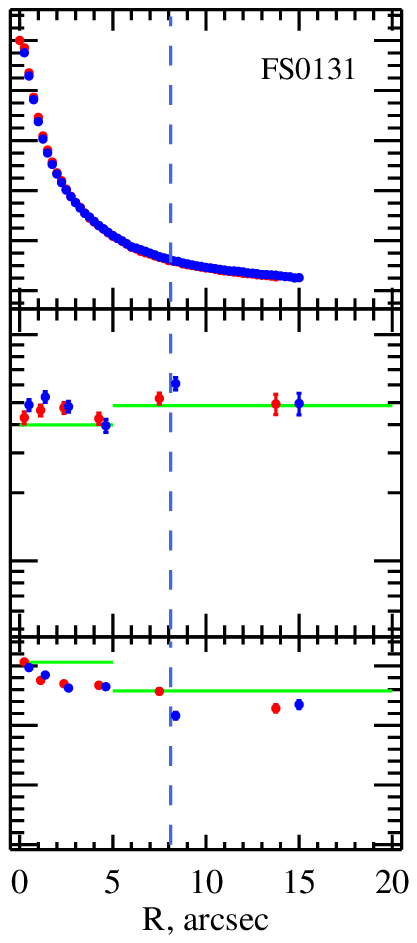}
\includegraphics[height=0.33\textheight]{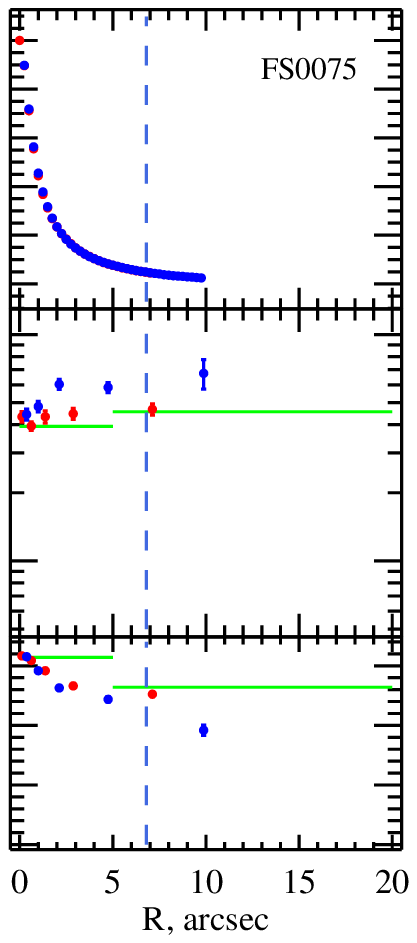}
\includegraphics[height=0.33\textheight]{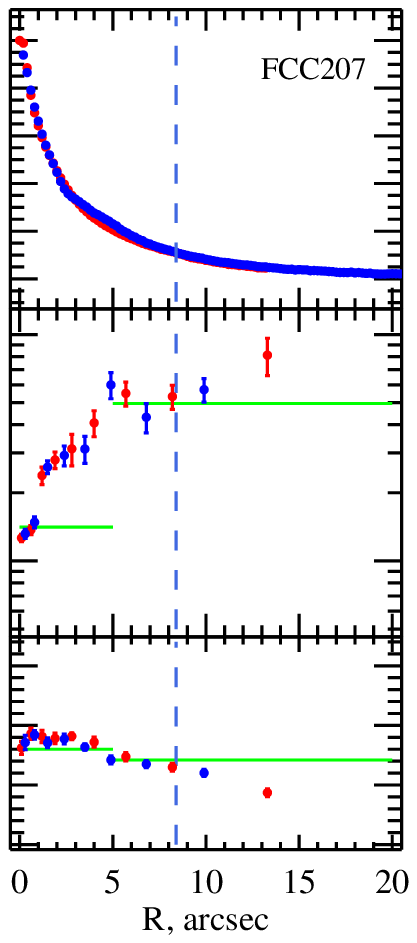}
\includegraphics[height=0.33\textheight]{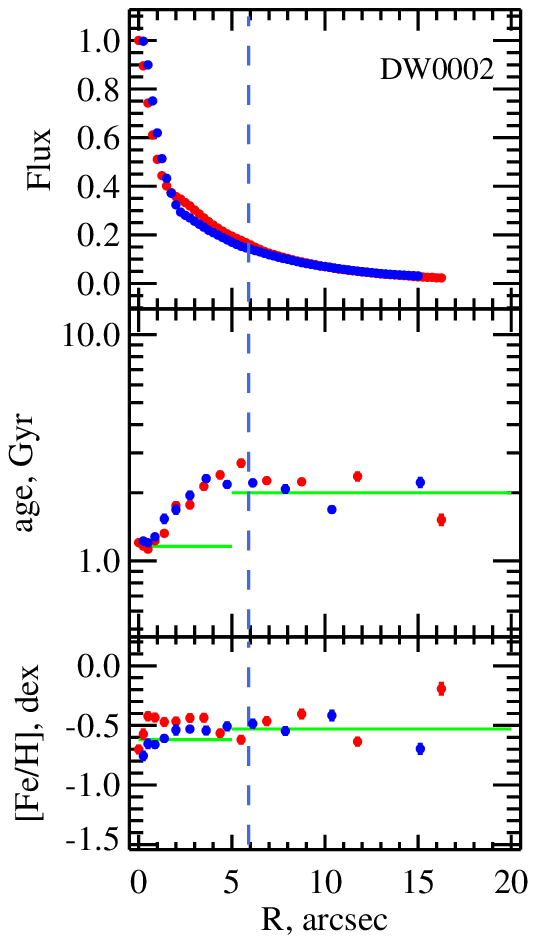}
\includegraphics[height=0.33\textheight]{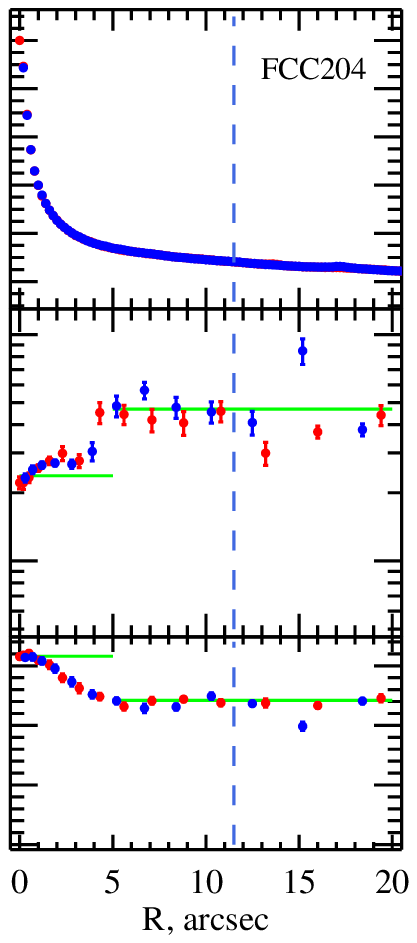}
\includegraphics[height=0.33\textheight]{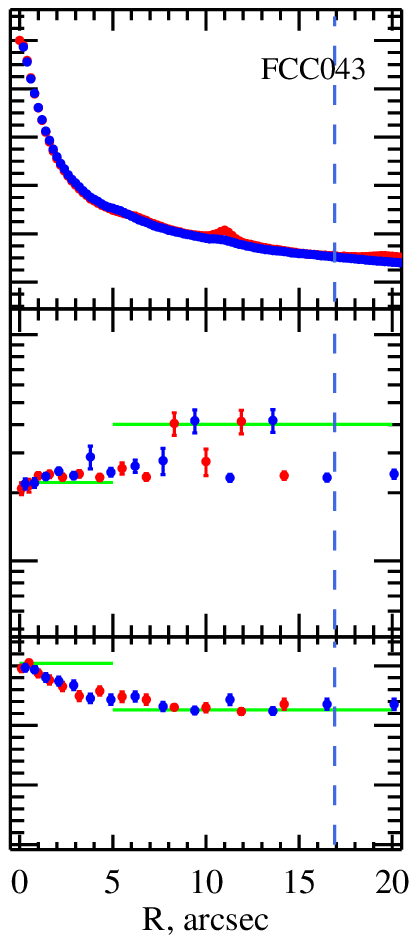}
\includegraphics[height=0.33\textheight]{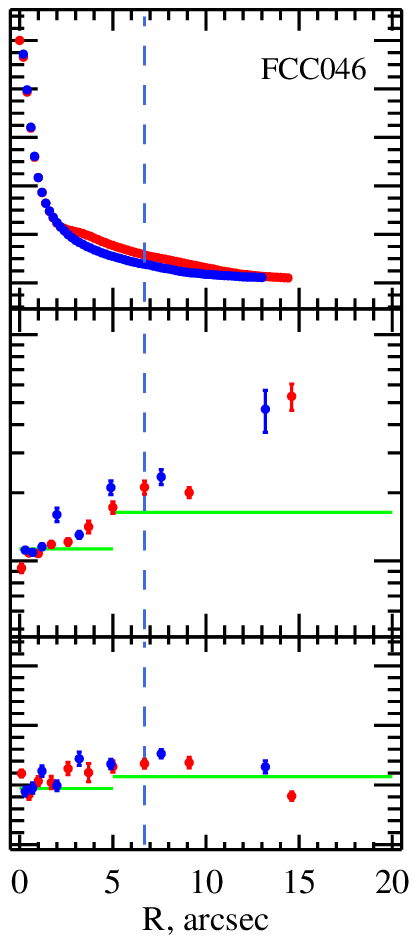}
\caption{ Radial profiles of the dEs with the longest formation time
  scale (see paragraph \ref{subsec:sfh}). The symbols are as in
  Fig.~\ref{fig:radprof1}.  }
\label{fig:radprof2}
\end{figure*}

\subsection{Analysis method}

The principle of our analysis is to compare an observed spectrum with
a model of a population broadened to account for the internal
kinematics. A least-square minimization provides the parameters of the
population model, age and metallicity in the case of a single stellar
population (SSP).  The main characteristics of the method, described
and validated in \citet{koleva08a}, is to fit the full spectrum,
pixel-by-pixel, while the most common approach uses spectrophotometric
indices.  The SSP-equivalent ages and metallicity for central
extractions of the present data were found perfectly consistent with
Lick indices and three times more precise \citep{michielsen08}.
Beside the precision, another advantage of the method is its
insensitivity to the presence of emission lines, that can be either
masked or fitted with additive Gaussians.

We are using population models generated with Pegase.HR
\citep{PEGASEHR} assuming Salpeter IMF and Padova isochrones and build
with the ELODIE.3.1 library \citep{PS01, ELODIE31}.  These models have
a well calibrated line spread function (LSF, i. e. instrumental
broadening) of 0.55~{\AA} corresponding to a resolution R $\approx$
10000 or instrumental velocity dispersion $\sigma_{\rm instr} = 13$
{\kms} at $\lambda = 5500$~{\AA}.  

The ELODIE library, as any other empirical libraries, consists mostly 
of stars of the solar neighborhood and therefore presents the abundance 
pattern of this environment. The chemical composition of the stars is 
scaled-solar at high metallicity, but becomes enhanced in $\alpha$-elements
at low metallicity ([Fe/H] $<$ -1). A consequence of this feature is that
low metallicity globular clusters ([Fe/H] $\approx$ -2; [Mg/Fe] $\approx$ 0.3) 
are correctly fitted, while for high metallicity clusters 
([Fe/H] $\approx$ -0.5; [Mg/Fe] $\approx$ 0.3) the models will under-fit 
the Mg lines. This effect is illustrated in \citet{prugniel07}.
Models of Lick indices with various $\alpha$-elements abundances
\citep{thomas03} are commonly used to measure the  $\alpha$-enhancement,
and the equivalent
has been also explored for spectral models \citep{koleva08e}:
The empirical models were corrected for the differential
effect of the  $\alpha$-enhanced computed in a theoretical library
\citep{coelho05}. This method proved to be efficient to measure
the enhancement of globular clusters and to absorb the
residuals on the Mg$_b$ feature. It was also found from Monte-Carlo
simulations that the measurements of [Fe/H] and [Mg/Fe] with full-spectrum 
fitting are independent. In other words, a mismatch of [Mg/Fe] does not
affect the measurement of [Fe/H].

Previous studies found that the populations of dEs have close-to-solar
[Mg/Fe] \citep{thomas03b}, and in this luminosity range (similar to NGC~205)
the metallicity is around [Fe/H] $\approx$ -0.5. Therefore, 
we do not expect an important Mg$_b$ residuals, but possibly
a small under-abundance with respect to the library.

As we know that this will not affect the [Fe/H] measurements,
we preferred to conservatively use the empirical library
instead of the $\alpha$-elements resolved semi-empirical one
which is not tested in details. Examination of the residuals of
the fits to the dE spectra does not reveal significant residuals on
the Mg lines (in most of the cases the residuals in the Mg$_b$ 
region are inside the 1$\sigma$ errors).

The analysis is made with the
\textsc{ulyss}\footnote{\url{http://ulyss.univ-lyon1.fr}} package
\citep{koleva08c}.
The first step is to reduce the resolution of the
models to that of the observations. To do this, we first determine the
relative LSF between the observations and the models as a function of
the wavelength, and then we convolve the models by the appropriate
function (for more details, see \citealt{koleva08a, koleva08c}).  The
LSF analysis was made using twilight spectra taken with the same
setup.
 
\subsection{Radial profiles}

The analysis was performed with a SSP model
parameterized by its age, metallicity, velocity and velocity
dispersion. A multiplicative polynomial of order 20 was included in
the model and the outliers were iteratively rejected. 
Each independent
1D-spectrum was generated by binning together spectra until S/N=20 was
reached, starting from the central spectrum. The last point has
S/N$>$10. 
The radial profiles for the 16 galaxies are shown in
Figs.~\ref{fig:radprof1} and \ref{fig:radprof2}.  For three galaxies, 
the H$_\beta$, H$_\gamma$ and [O{\sc iii}]$~\lambda5007${\AA} emission lines 
were fitted with gaussians in the same time as all the other free parameters.  
The error bars, estimated from the noise and quality of the fit, give the
independent errors on the age and on the metallicity and do not take
into account the age-metallicity degeneracy (a metallicity offset is
compensated by a bias of the age): some points are deviating from a
smooth profile by more than the error bar, but they lie on the same
age-metallicity degeneracy line (or actually age-metallicity-$\sigma$
surface). The error bars determined from Monte-Carlo simulations
are about twice larger, as they take also into account the degeneracies.

The first thing to notice is the
high degree of symmetry of the profiles (asymmetries seen in FCC043 and
266 are due to background galaxies along the slit, see Appendix).
Most galaxies show significant
metallicity gradients, with metallicity declining by 0.5~dex over a
distance of one half-light radius, on average, with some exceptions
discussed later in this paper.

\subsection{Inner 1~arcsec, corrected for line-of-sight contamination}

\begin{figure}
\includegraphics[width=0.5\textwidth]{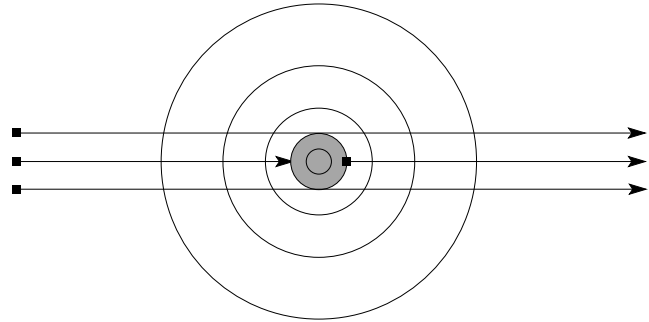}
\caption{The spectrum at a projected radius of 0.5~arcsec (the upper and
  lower lines of sight) is to a good approximation the amount light
  that is contributed by stars along the line of sight in front of and
  behind the inner 1~arcsec diameter sphere (the middle line of sight).}
\label{fig:extraction}
\end{figure}

We also study the inner 1~arcsec extraction, corrected for
contamination by stars along the line-of-sight that are outside of the
inner sphere with 1~arcsec diameter. The procedure is clarified in
Fig. \ref{fig:extraction}. The spectrum at a projected radius of 0.5~arcsec
is to a good approximation the amount light that is contributed by
stars along the line of sight in front of and behind the inner 1~arcsec
radius sphere. This means that a fraction
\begin{equation}
{\cal F}_{\rm outside~1~arcsec} = \frac{\sum_i F_{{\rm 1~arcsec},\lambda} r_i}{\sum_i F_{i,\lambda}r_i} \label{foutside}
\end{equation}
of the light in the 1~arcsec extraction comes from stars outside the
inner 1~arcsec diameter sphere. We fitted SSPs to the corrected
1~arcsec (Table~\ref{table:SSP}). In 6 cases, when the photometric
profile is flat, the correction implied a subtraction of a too large fraction
of the light, and no correct fit could be obtained.

In 7 cases out of 10, the population in the core is significantly younger
than in the 1~arcsec extraction, by 0.3 to 0.7 dex. The metallicities
do not show systematics. The most stricking case is FS~29 where the 
nucleus appears very young and metal-rich.

\begin{table*}
\centering
\begin{minipage}{180mm}
\caption{Results from the SSP fits of the central 1~arcsec, 1R$_e$ and the 1~arcsec corrected for the light on the line-of-sight,
and light-averaged age and metallicity determined from the star formation histories. The ages are in Gyr and metallicity in dex, relative to the Sun.}
\label{table:SSP}
\begin{tabular}{lcccccccccc}
\hline
&\multicolumn{6}{c}{SSP-equivalent}&\multicolumn{4}{c}{Computed from SFH}\\
&\multicolumn{2}{c}{1~arcsec}&\multicolumn{2}{c}{1R$_e$}&\multicolumn{2}{c}{1~arcsec corrected}&\multicolumn{2}{c}{1~arcsec}&\multicolumn{2}{c}{1R$_e$}\\
NameA  & age, Gyr & [Fe/H], dex& age, Gyr & [Fe/H], dex& age, Gyr & [Fe/H], dex& age& [Fe/H]& age& [Fe/H]\\
\hline
FCC136 &  5.03 $\pm$  0.55 & -0.25 $\pm$  0.03 &  8.58 $\pm$  0.71 & -0.59 $\pm$  0.03 &  3.64 $\pm$  2.33 & -0.45 $\pm$  0.18 & 7.0 & -0.26 & 8.8 &-0.60\\
FCC266 &  3.85 $\pm$  0.30 & -0.48 $\pm$  0.03 &  7.04 $\pm$  0.40 & -0.81 $\pm$  0.03 &  --               &  --               & 6.2 & -0.36 & 9.0 &-0.85\\
FCC150 &  3.84 $\pm$  0.19 & -0.39 $\pm$  0.02 &  5.45 $\pm$  0.27 & -0.61 $\pm$  0.03 &  1.83 $\pm$  2.23 & -0.32 $\pm$  0.19 & 6.5 & -0.45 & 6.3 &-0.67\\
FCC245 &  4.46 $\pm$  0.64 & -0.65 $\pm$  0.04 &  6.30 $\pm$  0.55 & -0.65 $\pm$  0.04 &  2.84 $\pm$  2.29 & -0.76 $\pm$  0.30 & 5.6 & -0.68 & 8.5 &-0.70\\
DW1    &  3.74 $\pm$  0.16 & -0.23 $\pm$  0.01 &  4.82 $\pm$  0.10 & -0.26 $\pm$  0.01 &  1.78 $\pm$  1.31 & -0.15 $\pm$  0.20 & 5.3 & -0.25 & 7.1 &-0.27\\
FS29   &  2.92 $\pm$  0.05 &  0.01 $\pm$  0.01 &  4.25 $\pm$  0.08 & -0.19 $\pm$  0.01 &  0.70 $\pm$  0.54 &  0.46 $\pm$  0.13 & 4.3 & -0.00 & 5.7 &-0.21\\
FS373  &  1.55 $\pm$  0.02 &  0.08 $\pm$  0.01 &  4.01 $\pm$  0.08 & -0.39 $\pm$  0.01 &  --               &  --               & 2.3 &  0.05 & 5.8 &-0.38\\
FS76   &  6.04 $\pm$  0.07 &  0.14 $\pm$  0.01 &  6.92 $\pm$  0.21 & -0.08 $\pm$  0.01 &  --               &  --               & 6.2 &  0.13 & 6.9 &-0.06\\
FCC288 &  4.08 $\pm$  0.50 & -0.58 $\pm$  0.03 &  3.85 $\pm$  0.12 & -0.44 $\pm$  0.01 &  --               &  --               & 6.0 & -0.65 & 4.5 &-0.55\\
FS131  &  3.99 $\pm$  0.19 &  0.03 $\pm$  0.01 &  4.58 $\pm$  0.19 & -0.20 $\pm$  0.02 &  --               &  --               & 4.2 &  0.02 & 6.5 &-0.22\\
FS75   &  3.93 $\pm$  0.09 &  0.07 $\pm$  0.01 &  4.81 $\pm$  0.15 & -0.19 $\pm$  0.01 &  4.51 $\pm$  1.23 &  0.16 $\pm$  0.05 & 4.1 &  0.09 & 7.0 &-0.23\\
FCC207 &  1.41 $\pm$  0.07 & -0.70 $\pm$  0.05 &  4.64 $\pm$  0.15 & -0.77 $\pm$  0.02 &  0.34 $\pm$  1.78 & -0.68 $\pm$  0.96 & 1.8 & -0.68 & 5.0 &-0.86\\
DW2    &  1.16 $\pm$  0.01 & -0.62 $\pm$  0.03 &  2.08 $\pm$  0.02 & -0.56 $\pm$  0.01 &  0.78 $\pm$  0.28 & -1.10 $\pm$  0.16 & 0.7 & -0.44 & 3.0 &-0.69\\
FCC204 &  2.38 $\pm$  0.08 &  0.08 $\pm$  0.02 &  4.68 $\pm$  0.13 & -0.29 $\pm$  0.01 &  1.50 $\pm$  0.44 &  0.15 $\pm$  0.07 & 2.9 &  0.06 & 5.2 &-0.43\\
FCC043 &  2.22 $\pm$  0.07 & -0.02 $\pm$  0.02 &  4.01 $\pm$  0.08 & -0.38 $\pm$  0.01 &  --               &  --               & 2.2 & -0.01 & 4.2 &-0.47\\
FCC046 &  1.13 $\pm$  0.03 & -1.03 $\pm$  0.04 &  1.65 $\pm$  0.05 & -0.93 $\pm$  0.04 &  1.17 $\pm$  0.26 & -1.31 $\pm$  0.47 & 0.2 & -0.43 & 1.8 &-0.72\\
\hline
\end{tabular}
\end{minipage}
\end{table*}

\section{Spatial distribution of the star-formation histories} \label{sec:sfh}

The stellar populations of dEs are commonly described by
SSP-equivalent ages and metallicities, derived from spectra
\citep{ggm03, michielsen08}. If the population is not co-eval, this
is essentially representing the 'light-dominant' epoch of star
formation and therefore the derived age and metallicity are often
called 'light-weighted' (though this short-cut is not rigorous).

Owing the high quality of the present spectra, we attempt here to
reconstruct more realistic star-formation histories (SFHs).

So far, SFHs of early-type dwarf galaxies
have been reconstructed for objects in the Local Group. Thanks to
their proximity, high-quality stellar color-magnitude diagrams (CMDs)
can be assembled and analysed by fitting synthetic CDMs to the
observed ones \citep{grebel00,dolphin02,dolphin03,cole07}. Beside the
significant old stellar population (age$>8$~Gyr) and absence of ongoing star formation, 
the diversity of recovered SFHs is startling. Some dwarf spheroidals are
genuine old stellar systems, with little or no star formation after
the initial burst that formed the system (\textrm{e.g}. Ursa Minor, Draco,
Sculptor). Others have been forming stars at an almost constant rate
until a few Gyr ago (\textrm{e.g.} Carina, Sagittarius, Leo~{\sc ii}). Leo~{\sc
  i} is an example of galaxy with a recent major star formation event,
between 3 and 1 Gyr ago, after which the star formation rapidly peters
out.

\begin{figure*}
\includegraphics[width=0.95\textwidth]{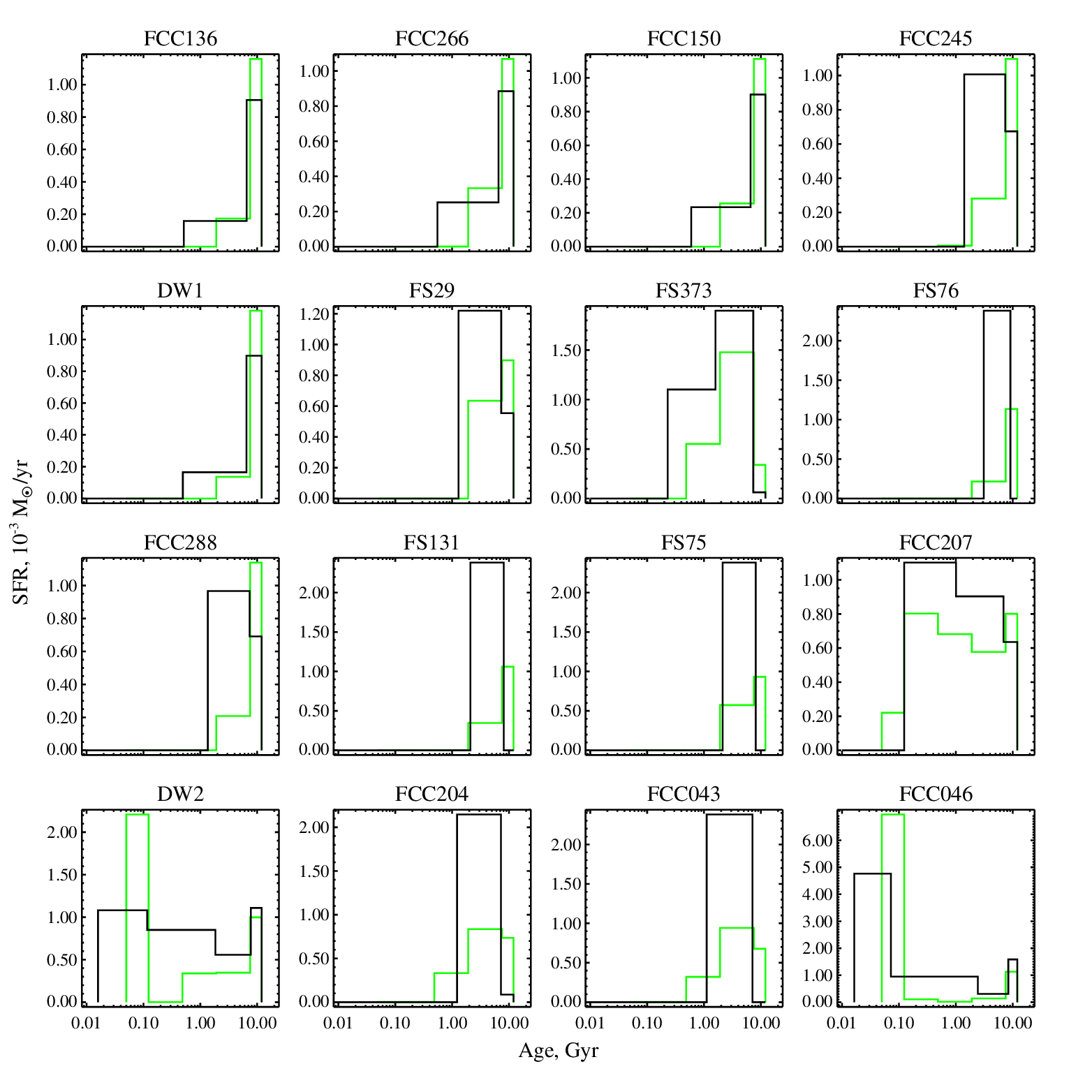}
\caption{ SFR (black for \textsc{steckmap}, green for \textsc{ulyss}) for the inner arcsec 
of the dEs (with assumed mass of 10$^7$	\textrm{M}$_{\odot}$ .)
}
\label{fig:sfr_1arc}
\end{figure*}

Further than the Local Group the SFH has been determined from integrated light
spectra by decomposing an observation onto a SSP basis of three SSPs
\citep{lt06} (1-10 Myr, 10-500 Myr and 5 Gyr) or parametrized by a 
exponentially declining SFR (\citealp{pasquali05}; 
in the field dwarf galaxy APPLES~1).
In our sample, three galaxies (FCC46 and 207, NGC5898~DW2) show H$_\beta$ emission revealing the 
existence of a weak contemporary star formation activity.
(The [O{\sc iii}]~5007\AA{} emission lines discovered in the fits residuals of other galaxies
of the sample: FCC150 and 204, NGC5898~DW1, FS76, 131 and 373, in both
the central and half-light radius extractions,
are probably due to evolved stars -- planetary nebulae and SN remnants).

The diversity of SFHs in the Local Group urges for a generalization
of time-resolved analyses. 

We will restore the SFH using two independent methods: \textsc{ulyss}
and  \textsc{steckmap}\footnote{
\url{http://astro.u-strasbg.fr/Obs/GALAXIES/stecmap_eng.html}} \citep{ocv06} 
that we will apply to both the central and the effective radius regions.

\subsection{Reconstruction of the star-formation histories} \label{subsec:sfh}

We will use two inversion methods: (i) a decomposition in 3 or 4 SSPs of free
age and metallicity using \textsc{ulyss} and (ii) a regularized decomposition on a 
basis of SSPs with \textsc{steckmap}.
 
\begin{figure*}
\includegraphics[width=0.95\textwidth]{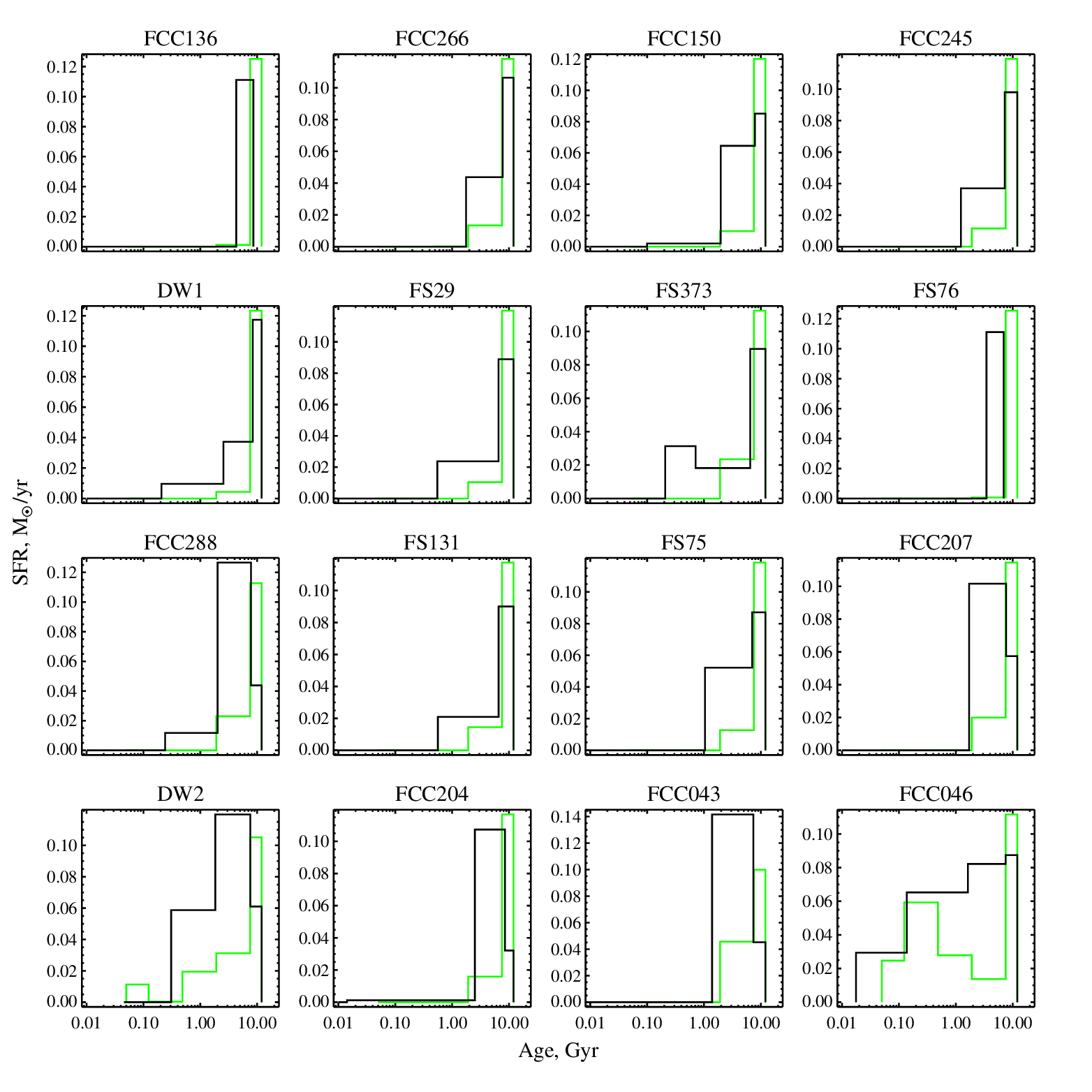}
\caption{SFR (black for \textsc{steckmap}, green for \textsc{ulyss}) for the inner 1R$_e$ 
of the dEs (with assumed mass of 10$^9$\textrm{M$_{\odot}$}).
}
\label{fig:sfr_1re}
\end{figure*}

With \textsc{ulyss} we can fit the observation against a positive linear combination of an
arbitrary number of sub-populations, setting some constraints on the age and
metallicity of each of them. The advantage of the method is its flexibility and
possibility to visualize the parameters space using \csq{} maps along the different
projections. The significance of a decomposition can be checked using Monte-Carlo 
simulations.

As a first approximation to a complex SFH we fitted two burst models, 
imposing the young population to be younger than 1 Gyr. Often, one of
the two components was pushing on this 1 Gyr age limit, indicating that the 
other box dominates the light and can probably be resolved at a higher time
resolution.

This drove us to approximate the SFH by three SSPs. 
In order to maintain the degree of freedom at the minimum,
we initially fixed the age of the oldest burst to 12~Gyr.
And we chose the acceptable limits of the two other
bursts as:
\begin{enumerate}
\item Young population aged between 12 and 800 Myr and with
  metallicities between -1.0 and 0.69 dex (note though that the SSP models
  are probably not very reliable for ages below 50 Myr and too high metallicities);
\item Intermediate population aged between 0.8 and 5 Gyr and 
with metallicities between -1.0 and 0.4 dex.
\end{enumerate}
The old population has a metallicity bounded between -2.3 and -0.4~dex.

This 3-populations fit is in general satisfactory in the sense that
the $\chi^2$ is significantly lower than for the SSP fit and that the
episodes are well within their assigned bounds in ages and metallicity.

\begin{figure*}
\includegraphics[width=0.9\textwidth]{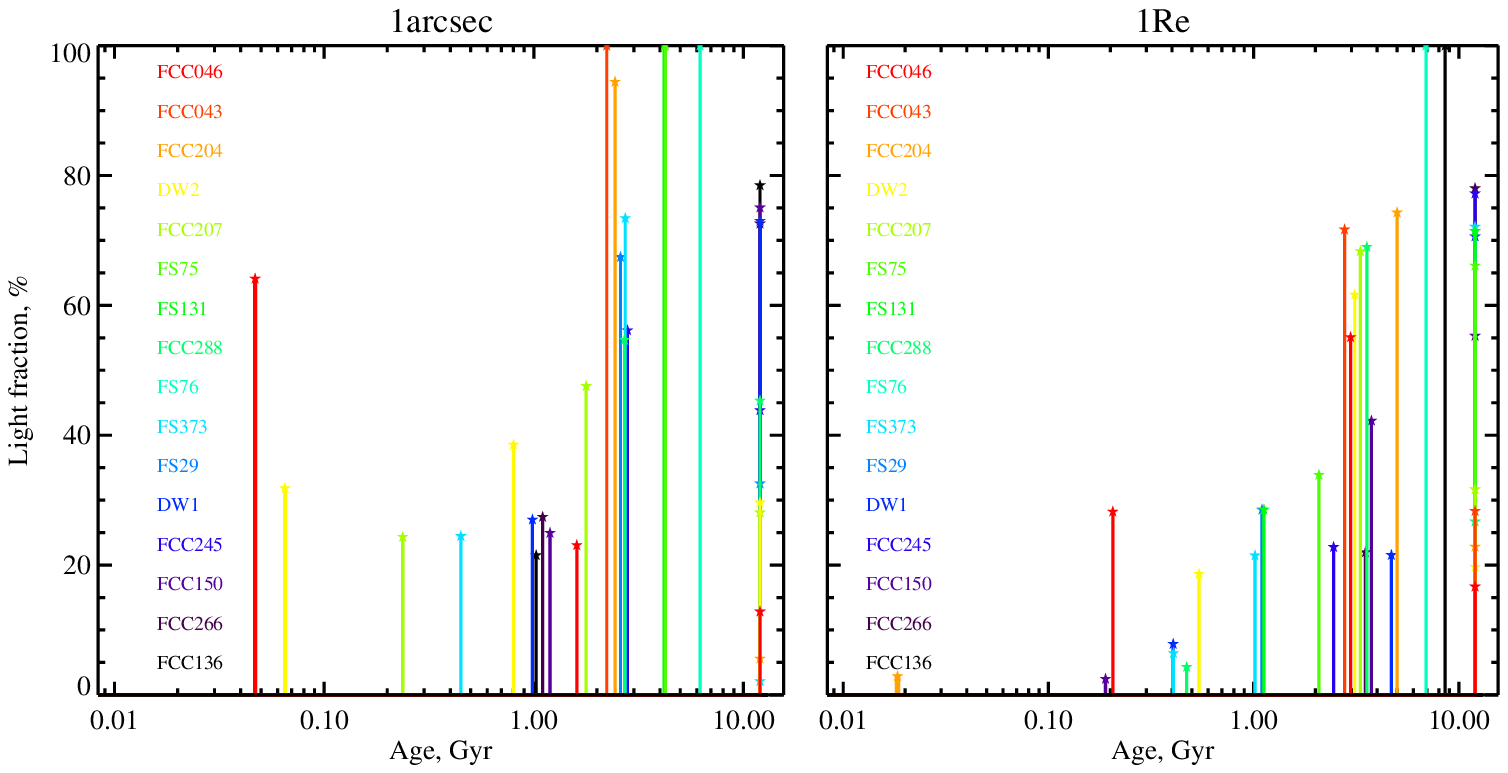}
\caption{Light fractions of each burst for all the galaxies.
Each galaxy is represented with a different color indicated in
the frame. Each star formation burst is shown as a vertical
bar whose position is its age and height its light contribution
to the total spectrum.}
\label{fig:light}
\end{figure*}

In some cases however one component appears to be below the 3-$\sigma$
detection limit or to reach an age-limit of the box. In the first
case we exclude this component and in the second we merge it with the
neighbouring one.  The results are in Table.~\ref{table:sfh} and the
distributions of the light fractions in the different components for
the two aperture extractions are shown on Fig.~\ref{fig:light}.  
For two galaxies, FCC046 and NGC5898~DW2, where a very young population
is detected, we fitted 4 bursts to resolve the lowest age bin. The
results are in Table~\ref{table:sfh4}.

\begin{table*}
\vbox to220mm{\vfil 
\vspace*{22.5cm}
\special{hscale=95 vscale=95 hsize=500 vsize=1500
hoffset=-89 voffset=-0 angle=0 psfile="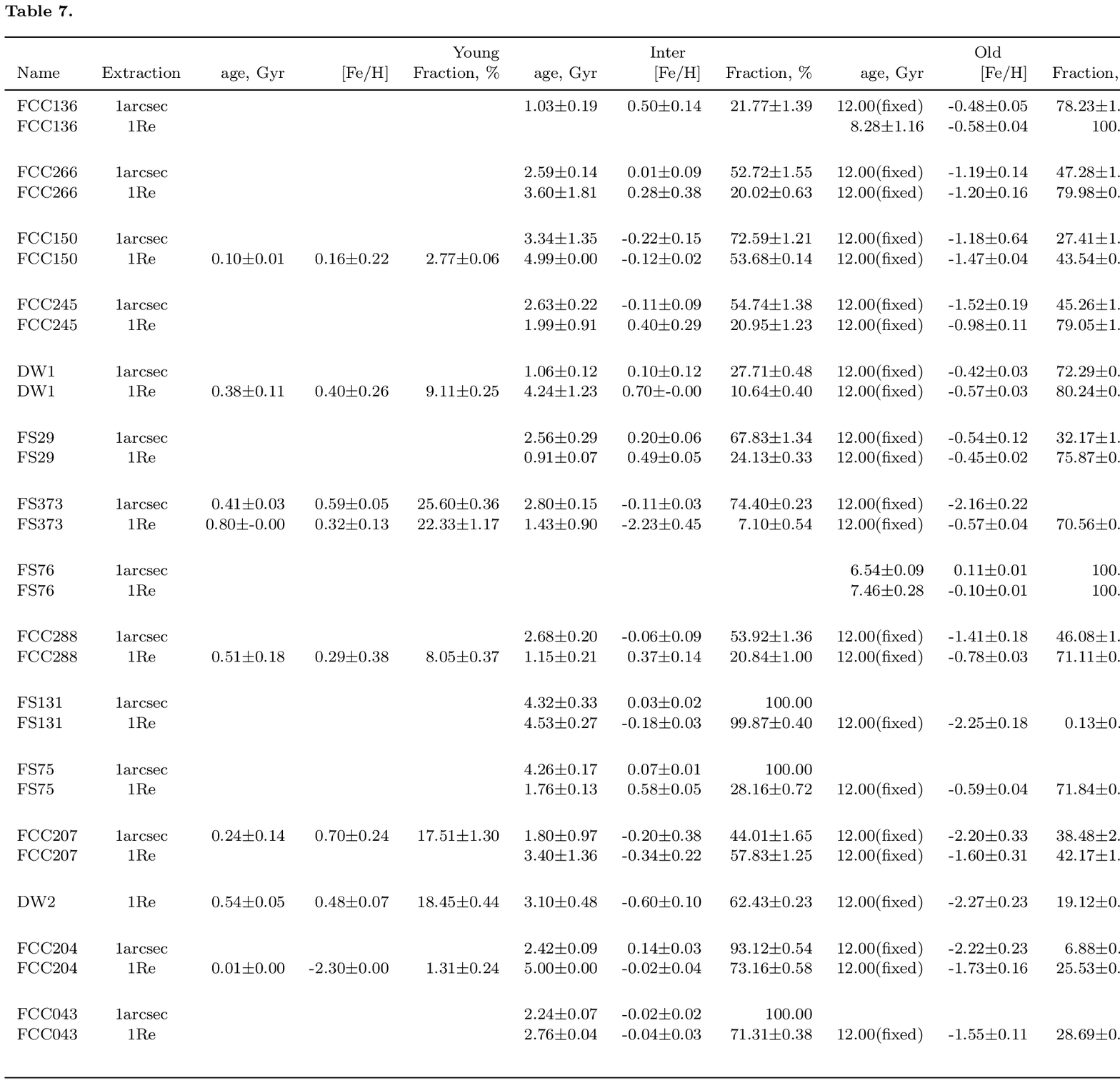"}
\caption{Population histories of the dEs, reconstructed from three episodes 
of star formation. In the first column is the name of the galaxy and 
its extractions, the second is for the \csq-s, followed by the 
age (in Gyr), [Fe/H] (in dex) and light fraction
for the young (from 0-0.8Gyr), intermediate (from 0.8 to 5~Gyr) and
old ($\sim$12~Gyr) populations.
} 
\label{table:sfh}
\vfil}
\end{table*}

We detect an important old component (over 10 Gyr) in all galaxies except
FS~76 (in FS75 and 131 the old population is detected only in the
outer extraction) , and a young population ($<$ 1 Gyr) in 8 galaxies:
FCC150, DW1, FS373, FCC288, FCC207, NGC5898~DW2, FCC204, FCC46.  The young
population is generally more concentrated than the oldest components
(Fig.~\ref{fig:light}).

To investigate the relation between the SFH and the other
characteristics, we ordered the galaxies according to the
mass fraction contained in the old population (averaged between
the two extractions), as it gives a idea of the formation
time-scale. Tables~\ref{table:1}, \ref{table:SSP}, \ref{table:sfh} 
and Fig. \ref{fig:radprof1}, \ref{fig:radprof2}, \ref{fig:sfr_1arc} 
and \ref{fig:sfr_1re} are sorted in this order.

\begin{table}
 \centering
 \begin{minipage}{\columnwidth}
  \caption{Four-epochs star formation history for the two 'young' galaxies FCC046 and NGC5898~DW2.
The light is the light fraction in per cent, ages are in Gyr and metallicty in dex, relative to the Sun.
    \label{table:sfh4}}
  \begin{tabular}{llcccc}
  \hline
\multicolumn{2}{c}{Burst} &\multicolumn{2}{c}{FCC046}& \multicolumn{1}{c}{NGC5898~DW2}  \\
&&\multicolumn{1}{c}{1 arcsec}&\multicolumn{1}{c}{1 \Reff}&\multicolumn{1}{c}{1 arcsec}\\
  \hline
\multirow{3}{*}{1} & light & 32$\pm$1       & 5$\pm$1      &17.5$\pm$0.5  \\
                   & Age   & 0.016$\pm$0.001& 0.02$\pm$0.02&0.015$\pm$0.03\\
                   & [Fe/H]& 0.00$\pm$0.17  & 0.0$\pm$0.5  &-0.8$\pm$0.3  \\ 
\smallskip \\
\multirow{3}{*}{2} & light & 40$\pm$1       & 29$\pm$1     &38$\pm$1      \\
                   & Age   & 0.13$\pm$0.03  & 0.26$\pm$0.08&0.22$\pm$0.02 \\
                   & [Fe/H]& 0.13$\pm$0.13  & 0.3$\pm$0.2  &0.44$\pm$0.05 \\ 
\smallskip \\
\multirow{3}{*}{3}  & light& 7.5$\pm$1.2    & 43$\pm$2     &18$\pm$1      \\
                   & Age   & 4.7$\pm$2.0    & 2.9$\pm$1.4  &3.5$\pm$0.8   \\
                   & [Fe/H]& -0.7$\pm$0.5   & -0.8$\pm$0.3 &-0.1$\pm$0.2  \\ 
\smallskip \\
\multirow{3}{*}{4}  & light& 20$\pm$2       & 22$\pm$2     &27$\pm$1      \\
                   & Age   & 12             & 12           &12            \\
                   & [Fe/H]& -2.3$\pm$0.4   & -1.8$\pm$0.4 &-1.7$\pm$0.1  \\
 \hline
\end{tabular}
\end{minipage}
\end{table}

This SFH reconstruction with \textsc{ulyss} allowed to test step-by-step
different hypotheses (choice of the number of bursts and of their age
distribution, significance of each burst) and we gained confidence in
the relevance of the results. But seeking for a more objective and
automatic method, we face the known difficulty of the instability of
this inversion.  Several methods were devised to cope with this
situation:~\citet{moultaka05},
\textsc{starlight}\footnote{\url{http://www.starlight.ufsc.br/}}
\citep{STARLIGHT}, \textsc{moped} \citep{MOPED}, \textsc{steckmap} and more recently
\textsc{vespa} \citep{VESTA}.

We analysed the sample with \textsc{steckmap}, using the same population models
(Pegase.HR with Elodie 3.1) and choosing a resolution into 5 age bins.
Any number of age bins can be required with \textsc{steckmap}, and the program
determines the SFH with the a priori that it should be smooth. As we
have shown that our 3 or 4 SSPs decomposition is realistic, we chose
to use \textsc{steckmap} with 5 age boxes logarithmically spaced. These 
fixed-limits boxes give a resolution comparable to the \textsc{ulyss} fits.

The SFR is represented in Fig.~\ref{fig:sfr_1arc} and \ref{fig:sfr_1re} 
as a function of time for
both the \textsc{ulyss} (in black) and \textsc{steckmap} (in green) fits for the 1
arcsec and 1 Re extractions. The spectral analysis does not provide
immediately the SFR, but rather give the optimal weight of the
different SSP components. To derive the SFR we have to assume that
these individual SSPs approximate in fact the SFH over a period of
time, and the simplest is to consider that the SFR is constant in each
box. Then one has to choose the limits of these boxes.  Setting the
limits at medium location between the ages (or more properly their
logarithm) of the components, and deriving the outer limits such that
the age of the extreme components sit in the middle of their boxes, is
the first idea. However, one has to realize that the size of the older
box, which is the longer, is not really constrained by the data. The
only physical constraint is that the initial burst lasted at least one
Gyr, otherwise [Mg/Fe] would have been enhanced, but we we are free to
consider a width between a couple of Gyr and about 6 Gyr (from 8 to 14
Gyr). This implies an uncertainty by a factor 3 on the initial
SFR. We chose to fix the beginning of the star-formation at 14 Gyr in
the past, therefore the reported initial SFR is a lower limit of the
mean SFR during the first Gyr of the galaxy. It is likely that during
this period the SFR passed through a succession of more violent
episodes. The choice of the lower limit of the most recent box is less
critical, and because of the exponential decrease of the luminosity of
a population, we set the limit of the first box to the age of the
first SSP.

These assumptions have to be taken into account when reading 
Fig.~\ref{fig:sfr_1arc} and ~\ref{fig:sfr_1re}, but at least these 
figures allow a direct comparison of the
histories reconstructed with both methods. 

\begin{table*}
\vbox to220mm{\vfil 
\vspace*{22.5cm}
\special{hscale=95 vscale=95 hsize=500 vsize=1500
hoffset=-89 voffset=-0 angle=0 psfile="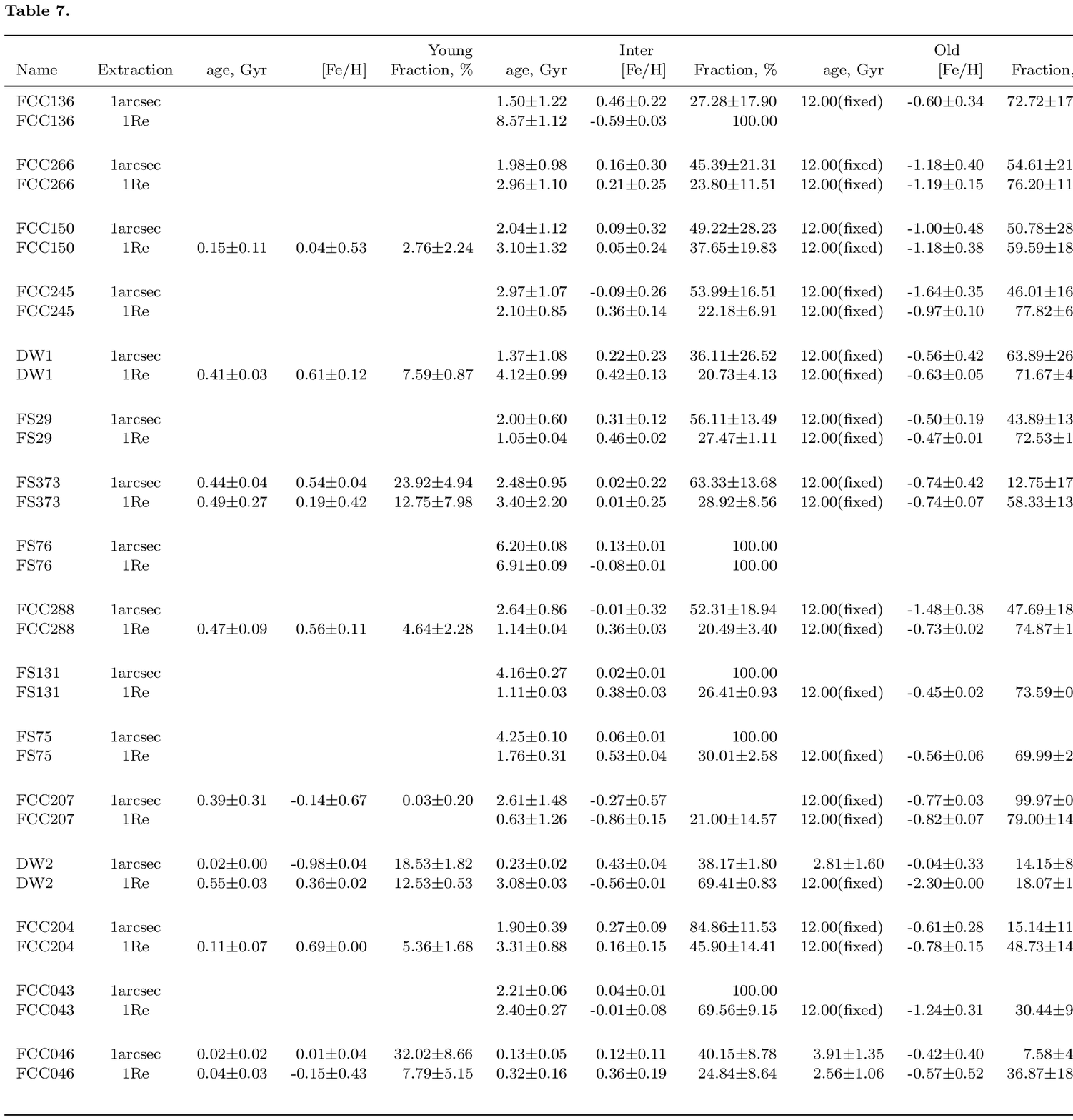"}
\caption{Population histories of the dEs, from Monte-Carlo simulations.
Same caption as Table~\ref{table:sfh} .
} 
\label{table:mc}
\vfil}
\end{table*}

\begin{figure}
\includegraphics[height=0.33\textheight, width=\columnwidth]{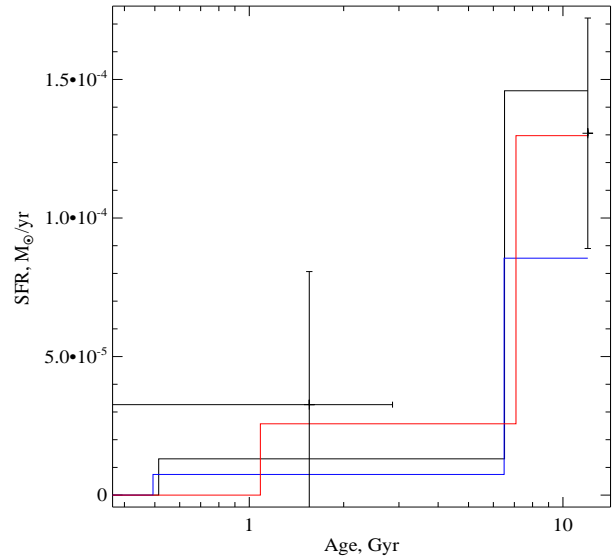}
\caption{Reliability tests of the SFH of the inner one arcsecond in FCC136 
taken as an example: SFR vs. time recovered 
with  \textsc{ulyss}.
The black line is the reference SFH determined with Pegase.HR/Elodie3.1 
models using Salpeter IMF (direct solution). The crosses are the Monte-Carlo 
solution. The red line represents the solution with Vazdekis/Miles grid of models, 
and the blue with Pegase.HR/Elodie3.1/Kroupa IMF models.
}
\label{fig:sfhvalid}
\end{figure}

\subsection{Reliability of the determined star formation histories}

Constraining the star formation histories is a considerably
harder task than fitting SSPs, and it is legitimate to question
the validity of the solutions. Adding degrees of freedom
by adjusting a combination of SSPs is likely to reduce the \csq,
but this does not warrant the correctness of the decomposition.

Multiple local \csq minima may appear, and they may not be significantly 
distinguishable, in the sense that changing
the noise realization, the wavelength range or some systematic
defects in the model or in the observation, may lead to another
solution. If these various solutions are close this will not
question the validity of the reconstruction, but it may not
be the case. 
Some solutions may be a posteriori rejected on the
basis of physical arguments, but rather than adding this
additional layer, the inversion programs regularize the solution
by adding constraints. \textsc{steckmap} imposes a smooth relation
between the metallicity and the age of the components. With
\textsc{ulyss} we decomposed the history in a small number
of epochs represented by SSPs (3 or 4 bursts). 

The consistency between the two approaches (See Figs.~\ref{fig:sfr_1arc}
and \ref{fig:sfr_1re}) gives some confidence in the results. 
And to go further, we are testing below the robustness of the
solution with respect to the noise with the help of 
Monte-Carlo simulations, and we are considering other physical
causes that may alter the solution : (i) the sensitivity
to the population model, (ii) the the IMF and (iii) the presence
of a blue horizontal branch.

\subsubsection{Stability with respect to the noise}

One of the basic tests is to make Monte-Carlo simulations with
an added noise equivalent to the estimated noise. In the present
case, the noise is considered gaussian, with a dependence
between the pixels due to the rebinning.

We performed 200 simulations for each galaxy and both extractions.
Each time a different random noise realisation is added and the inversion
is made. Table\,\ref{table:mc} gives the average and the dispersion of
the measured parameters; they can be compared with the results
of the direct fit given in Table\,\ref{table:sfh}.

While the error bars of the direct fit ignore the degeneracies
between the parameters, the Monte-Carlo simulations take them
into account. The most affected from degeneracies parameter is
the fraction, while the direct fit gives a precision of 1 per cent,
the Monte Carlo simulations show much less confidence 
($\approx$\,20 per cent). For the other parameters, this effect of booming
the errors is smaller, $\approx$\,5 times for the ages (from
few hundreds Myr up to 1\,Gyr) and around two times for the metallicities. 
The Monte-Carlo estimated values of the parameters are
consistent with the direct fit within the error bars.

As an example, in Fig.~\ref{fig:sfhvalid} the Monte-Carlo solution is plotted over the
direct solution for FCC~136 in the 1 arcsec extraction.

\subsubsection{Sensitivity to the population model}

\citet{koleva08a} compared SSPs models from Pegase.HR/Elodie
with models from Vazdekis built with the Miles library\footnote{
\url{http://www.iac.es/galeria/vazdekis/vazdekis_models_ssp_seds.html}} 
(VazMiles). They found a very satisfactory agreement,
despite the fact that the two models use independent ingredients.

We repeated the analysis using VazMiles instead of Pegase.HR/Elodie.
The result for FCC~207 is shown on Fig.~\ref{fig:sfhvalid}. 
The agreement with the original fit is satisfactory.

\subsubsection{Sensitivity to the IMF}

The initial mass function of the stars is one of the ingredients
of the population model that is not constrained by the observations.
We may wonder if it is possible that the detection of an important
old (12 Gyr) component, though physically reasonable, 
is not biased by the assumed IMF. It is probably possible to 
build an ad hoc IMF to mimic the old component, but this would
lack physical support.

In Fig..~\ref{fig:sfhvalid} we are showing the effect of changing the
adopted Salpeter's IMF for Kroupa's \citep{kru93}. 
Though the choice of the IMF may have a dramatic 
effect on the mass-to-light ratio, the considered alternative
does not affect the SFH.

\subsubsection{Sensitivity to a blue horizontal branch}

An important effect can be due to extended blue
horizontal branches or to blue stragglers which are not represented
in the models. \citet{koleva08a} have shown that this leads to 
underestimated ages or mimics a young sub-population in
Galactic globular clusters, as also suggested by \citet{li08}.
In the Local Group, \citet{mapelli07} detected blue stragglers.

However, compared to the effect seen in \citet{koleva08a}, where
the excess of blue stars was seen as a small and well detached young
component, the SFH presented here are considerably more time resolved.
So, we believe that the possible presence of blue stragglers cannot be
an important explanation for the extended SFH.

\bigskip
\noindent
After this series of tests, we are reasonably confident in the
reliability of the reconstructed SFH.

\section{Discussion} \label{disc}

The main features standing out from our analysis are the large
diversity of the zoo of SFHs and of the radial gradients of
SSP-equivalent age and metallicity.

In this section we will discuss the radial gradients detected
in the populations (\ref{sect:gradient}) and the star formation history 
(\ref{sect:sfh}). Then we will have a close look at the central region
(\ref{sect:center}) and we will critically review the hypothesis
concerning the origin of the dEs (\ref{sect:origin}).

Our sample is too small to draw any firm conclusions about the 
connection between the environment and the population parameters.
However, these data do not suggests such a connection.
It naturally does not imply that the environment is not a key 
factor controlling the evolution.

\subsection{Radial gradients of the stellar populations} \label{sect:gradient}

Many of the observed dEs show a radially increasing age. The extremes
are formed by FCC288 (dS0(7),N), FS76 (dE1), FS131 (dE5,N), and
NGC5898~DW1 (dE3), which show almost no age gradient, on the one hand,
and FCC207 (dE2,N) and FCC046 (dE4,N) which show an age gradient of
over 5~Gyr between the centre and 2{\Reff}, on the other.

Most galaxies (10/16) also show significant metallicity gradients, with metallicity
declining by 0.5~dex over a distance of one half-light radius, on
average. This is comparable to what has been found for the Local Group
dE NGC~205 (-0.6$\pm$0.1 dex/\Reff, \citealp{koleva09}). The exceptions
to this pattern are discussed below.
We note that the age and metallicity opposite gradients conspire to
reduce the color gradients that may be observed.

\subsubsection{Flat metallicity profiles }

The galaxies standing out from the strong metallicity gradient rule
are FCC288 (dS0(7)), FCC204 (dS0(6),N), NGC5898~DW2 (dE6),
and FS29 (dE5), which, outside of the inner few arcseconds, have
essentially flat metallicity profiles out to the last data point. For
instance, FS29 and FCC204 have a metallicity gradient in the centre,
but the profile flattens out beyond 0.5~\Reff. The first four galaxies
in this list have the common property of being significantly flattened
and fast rotating. FCC288 and FCC204 contain clear evidence for
embedded stellar discs \citep{d03}. NGC5898~DW1 (dE3) and FCC046
(dE4,N) have flat metallicity profiles although they are somewhat
rounder objects. However, like FCC204 and FCC288, NGC5898~DW1 has a
positive $C_4$ coefficient which indicates that it has discy isophotes
(deviations of the isophotes from a pure elliptic shape are quantified
by expanding the intensity variation along an isophotal ellipse in a
fourth order Fourier series; $C_4$ measures the disciness/boxiness of
the isophotes, see \textrm{e.g.} \citet{d03}). Also, its flattening is much
stronger in the inner 10$''$ ($\epsilon \sim 0.5$) then in the outer
parts of the galaxy ($\epsilon \sim 0.2$). In the terminology of
\cite{lt07}, FCC204, FCC288, and NGC5898~DW1 would be classified as
dE(di) objects. Beside their morphology and flat metallicity profile,
the dE(di) galaxies do not share other properties. They span the whole
range of age, metallicity or formation time-scale of the sample and
are not found preferentially in either cluster or group environments.

The spatial chemical homogeneity in their stellar population
properties is reminiscent of the dwarf irregular galaxies (dIrrs): See
\textrm{e.g.} \cite{lsv06} and references therein for a discussion of the
non-detection of a radial gradient in the stellar and nebular
abundances measured in NGC6822 and in other dIrrs. There, supernova
feedback is expected to expell hot enriched gas which, after a delay
of over $10^8$~yr, rains back onto the galaxy's stellar disc
\citep{mafe99,riehe03}. This similarity between dIrrs and dE(di)s
could therefore be interpreted as another evidence for an evolutionary
link between these two types of dwarfs.

\subsubsection{Strong metallicity gradients}

Strong metallicity gradients are observed in 10 out of 16 dEs.  In
some Local Group dwarf spheroidals, such as the Sagittarius dwarf
\citep{alard01} and the Fornax dwarf \citep{bthi06} the gradients are
due to the imposition of populations of different ages and
metallicities.  The younger and more metallic populations being more
centrally concentrated, the combined population displays a negative
metallicity gradient and a positive age gradient.  According to this
formation scenario, one would expect the metallicity gradient to build
up over time. If the present sample can be considered as
representative of an homogeneous family whose variety is due to the
time elapsed since the last star formation event, we would expect to
see the strongest gradients in the objects with the youngest
SSP-equivalent ages. This effect is clearly not seen. This mostly says
that the sample is not a simple mono-variate family.

The metallicity gradients are also detected when we compare the 1
arcsec and 1~\Reff{} extractions, either through their SSP-equivalent
characteristics (Table~\ref{table:SSP}) or through their
light-averaged description computed from the SFH reconstruction given
in Table~\ref{table:sfh}.  In Table~\ref{table:gradient}, we give these
latter gradients and their values estimated after 3 Gyr of passive
evolution (aging of the stars, no kinematical mixing).  The gradients
are naturally on average smaller than measured on the profiles since
the light is integrated in a large aperture.  The average age gradient
from the observations is $0.26\pm0.23$ and after 3 Gyr vanishes to
$0.06\pm0.11$. The mean metallicity gradient evolves from
$-0.28\pm0.19$ to $-0.16\pm0.30$ (the dispersion can be reduced by
removing the objects with young populations for which the
characteristics of the old populations are more uncertain, and by
excluding the dE(di) galaxies discussed above.  This experiment,
though the important uncertainties, establishes that only 1/3 of the
observed mean gradient is due to the light contribution of the younger
and more concentrated population. This does not explain the whole
gradient which hence appear to be a genuine chemical abundance effect.

Such gradients metallicity are likely to occur if the star formation,
initially spread over all the galaxy, gradually becomes more centrally
concentrated, together with the increase of the metallicity of the gas.
This is shown in SPH simulations \citep{marcolini08,valcke08}.  The
self-regulation of the star formation by the feed-back results in
quasi-periodic bursts ('breathing' scenario,
\citealp{pelupessy04,stinson07}).  The fact that these gradients are
seen in the old population (mean gradient $0.3\pm0.4$) indicates that
this should have occurred during the few Gyr of the initial formation
when 60-90 per cent of the stellar mass formed.  The persistence of
these gradients over all the live of the galaxies implies that the
population has not been mixed, i. e. that the galaxy did not undergo a
violent relaxation (after a close encounter or tidal harassment) and
also sets constrains on the internal dynamics (the orbits do not mix
the population over a large range of radii).

\begin{table}
\centering
\begin{minipage}{\columnwidth}
\caption{Luminosity weighted age and Metallicity gradients.
Col. 1: log10( Age$_{{\rm 1 R}_e}$/Age$_{\rm 1~arcsec}$);
Col. 2: [Fe/H]$_{{\rm 1 R}_e}$ - [Fe/H]$_{\rm 1~arcsec}$;
Col. 3 \& 4: Same as 1 \& 2 after 3 Gyr of passive evolution; 
Col. 5 \& 6: SSP-equivalent gradients computed from Table~\ref{table:SSP}
The more uncertain values are noted with a colon.
}
\label{table:gradient}
\begin{tabular}{|l|c|c|c|c|c|c} \hline
&\multicolumn{2}{c}{Observed: T$_0$} & \multicolumn{2}{c}{T$_0$ + 3 Gyr} & \multicolumn{2}{c}{SSP-equivalent}\\
Name                & Age   & [Fe/H]&  Age   & [Fe/H]&  Age   & [Fe/H]\\
\hline
FCC136              &  0.11 & -0.35 & -0.06 & -0.20 &  0.23 & -0.34 \\
FCC266              &  0.19 & -0.52 &  0.03 & -0.45 &  0.26 & -0.33 \\
FCC150              &  0.32 & -0.61 &  0.05 & -0.38 &  0.15 & -0.21 \\
FCC245              &  0.18 & -0.02 &  0.09 &  0.05 &  0.15 &  0.00 \\
DW1$^a$    &  0.13 & -0.01 &  0.13 & -0.08 &  0.11 & -0.03 \\
FS029$^a$           &  0.11 & -0.20 &  0.05 & -0.21 &  0.16 & -0.20 \\
FS373               &  0.41 & -0.44 &  0.12 & -0.31 &  0.41 & -0.47 \\
FS076               &  0.05 & -0.19 &  0.03 & -0.19 &  0.06 & -0.22 \\
FCC288$^a$          & -0.12 &  0.09 & -0.11 &  0.10 & -0.03 &  0.14 \\
FS131               &  0.18 & -0.24 &  0.27 & -0.37 &  0.06 & -0.23 \\
FS075               &  0.23 & -0.32 &  0.24 & -0.47 &  0.09 & -0.26 \\
FCC207              &  0.44 & -0.18 &  0.02: &  0.40: &  0.52 & -0.07 \\
DW2        &  0.54 & -0.21 & -0.11: &  0.24: &  0.25 &  0.06 \\
FCC204$^a$          &  0.24 & -0.47 &  0.07 & -0.41 &  0.29 & -0.37 \\
FCC043              &  0.28 & -0.46 &  0.22 & -0.65 &  0.26 & -0.36 \\
FCC046\footnote{Flat metallicity profile dE(di) objects.} 
                    &  0.97 & -0.33 & -0.02: &  0.26: &  0.16 &  0.10 \\
\hline
\end{tabular}
\end{minipage}
\end{table}

\subsection{Star formation histories} \label{sect:sfh}

The detection of young population and even of ongoing star formation
is consistent with the findings of \citet{lt06} who found blue cores
in 4 to 5 per cent of a sample of about 450 dEs from the Virgo
cluster. We show here that the late star formation is concentrated
(Fig.~\ref{fig:light}) but still affects the bulk of the galaxy, giving
rise to age gradients (Fig.~\ref{fig:radprof1} \&
\ref{fig:radprof2}).  The mass fraction of the stars formed in the
last 5~6 Gyr is typically around 10~per cent.

Does this late star formation need the supply from a reservoir of
primordial gas, or can it be fueled by the winds of evolved stars as
it may be the case in NGC205 \citep{davidge05} or NGC185
\citep{martinez-delgado99}.  According to the population model that we
are using (Salpeter IMF), the old population returned 1/3 of its
stellar mass to the ISM. Obviously, for the galaxies of the present
sample that have constant or slowly decaying SFR (half of the sample),
the returned gas from the old stellar population is not a significant
contribution and we can exclude that these galaxies were stripped of
their gas at an old epoch.

As for the Local Group galaxies, the most stricking aspect is the
detection of an old population ($>$ 10 Gyr) in almost all the galaxies
of the sample. Only for FS76 the stellar population is consistent with
a single burst at an intermediate age (6-7 Gyr). In some cases, the
old population is not detected in the central extraction, but is seen
in the 1~\Reff{} extraction.  This old populations accounts for 15 to
80 per cent of the light and is the dominant fraction of the total
mass of stars that has been formed: 70-90~per cent.
The existence of populations contemporary of the old massive elliptical galaxies
was already infered from the
detection of an old star cluster in a dE galaxy by \citet{conselice06}.

This implies that, though the environment is probably the key for
quenching the star formation due to the stripping of the gas, its
triggering is likely an internal process. But it also tells that most
of the characteristics of the dEs where already set at an early epoch,
certainly near or above $z \approx 1$.  The diversity of observed SFH
may be accounted for by different quenching times, and a residual star
formation, fed by gas expelled by the evolved stars, may explain part
of the SFR observed in the last few Gyr.

The lack of young galaxies, i. e. without old populations, is consistent with
the non-detection of 'dark' galaxies by the blind HI survey ALFALFA
\citep{kent07}. Apparently, all gas clouds that could make stars, did it in
early epochs.

\subsection{Central stellar populations}  \label{sect:center}

When seen at high spatial resolution (e. g. from HST/ACS data) 
most of the dEs display a nucleus, i. e. a light excess over a 'reasonable'
extrapolation of the outer profile.  This nucleus may either be redder
or bluer than the surrounding population. The former case may be
explained by a higher metallicity, while the latter could either be
due to a younger age or to a lower metallicity
\citep{lotz04,cote06,lt06}. We will now discuss check if the presence
of a nuclei is reflected in the stellar population and we will examine 
the two galaxies where a 
kinematically decoupled core is known, and we will 
review the central profiles of the populations in our sample.

\subsubsection{Nucleated galaxies}

Half of the galaxies of our sample are nucleated: FCC~46, 150, 207 245 and 266,
FS~75 and 131 and NGC5898$\_$DW2. These galaxies are evenly distributed along
the sequence of star formation timescale. No systematics stand out for the 
examination of the star formation history (Fig.~\ref{fig:sfr_1arc})
and of the gradients (Fig.~\ref{fig:radprof1} \& \ref{fig:radprof2}). 
The four galaxies of the sample
with the lowest central metallicity are nucleated (FCC~46, 207, 245 and 
NGC5898$\_$DW2), and the correction of the line-of-sight contamination
enhances slightly the effect (see Table~\ref{table:SSP}).
But two of the other nucleated 
galaxies have a rather high metallicity in their centres (about solar).

Beside the fact that half of the nuclei coincide with
a low central metallicity, we cannot draw any conclusion concerning the
nature of the nuclei, or the relation between the presence of a nucleus and 
the properties of the galaxy.

\subsubsection{Galaxies with kinematically decoupled cores}

FS076 and 373 host kinematically decoupled cores (KDCs) \citep{d04}
observed within $\lesssim 1-2$~arcsec.  In FS373, the ellipticity
profile of the isophotes suggests the presence of a small central
stellar disc \citep{d04}.  Within the central 3~arcsec the
SSP-equivalent age raises regularly from 1.5 to 4.5 Gyr and the
metallicity decreases from 0.1 to 0.35 dex.  This behaviour is usual
in our sample. The SFH reconstruction Table~\ref{table:sfh} and
Fig.~\ref{fig:sfr_1arc} reveals the presence of a young and high
metallicity population contributing 25 per cent of the light in the
central extraction,  decreasing to 6 per cent in the 1 \Reff{}
aperture.  Beside the KDC, FS76 exhibits other peculiarities. Its
velocity dispersion, which initially rises with radius, starts to
decline sharply outside one half-light radius. Moreover, its surface
brightness profile is well described by a S\'ersic law with index
$n\approx 2$, making this galaxy more compact than most dEs of
comparable luminosity (M$_B=-16.7$~mag). All this has been interpreted
as evidence for a truncated dark halo and hence for the occurrence of
tidal stripping \citep{ddzh01}.  As FS373, it has a regular
metallicity profile decreasing from 0.2 to -0.2 dex within the central
3~arcsec and is consistent with a single burst about 6-7 Gyr ago in
the both extractions.

The presence of a KDC does not appear to be reflected by any 
pecularity of the stellar population. This indicates that the
KDC is not linked with a recent event that would have been 
marked by star formation, and should instead be a long-lived feature.
This supports the suggestion for a tidal origin discussed in \citet{d04}.

\subsubsection{Centrally depressed metallicity}

\begin{figure}
\includegraphics[width=0.98\columnwidth]{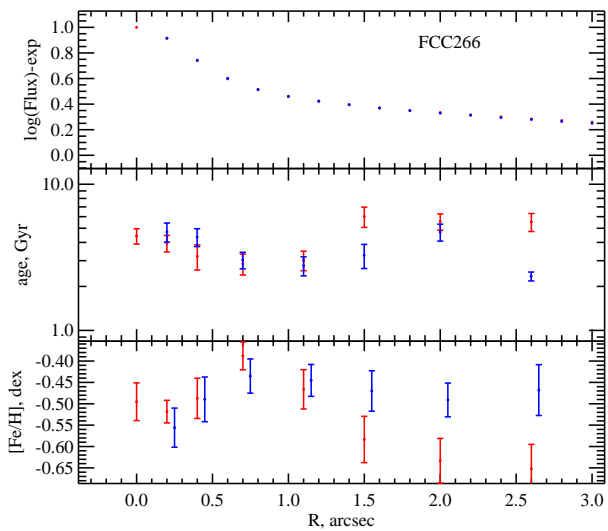}
\caption{Central SSP-equivalent profiles of FCC266.
The top panel is the excess over an exponential light profile fitted
using the long-slit spectra in the region between 2 and 5~arcsec.
The symbols are identical to Fig.~\ref{fig:radprof1}.
A finer radial binning than for Fig.~\ref{fig:radprof1} \& \ref{fig:radprof2}
was achieved by targetting a lower S/N.
}
\label{fig:depression}
\end{figure}

We have shown above that the general trend is negative
metallicity gradients at large scale. But in 5 cases
we detect an opposite trend in the central region.
Table~\ref{table:depression} lists the galaxies with centrally depressed 
metallicity,
with the magnitude of the effect and its radial extension.
The central profiles of age does not display any systematics.
As an illustration, Fig.~\ref{fig:depression} presents the profiles of FCC266.

\begin{table}
\centering
\begin{minipage}{\columnwidth}
\caption{Centrally depressed metallicities.
R$_{max}$ is the radius, in arcsec and in pc (not deconvolved for seeing 
effect), where the metallicity 
reaches its maximum.
$\Delta$[Fe/H] is the amplitude of the metallicity depression between its
maximum and the centre. For the last object of the table, FCC136, the detection
is marginal.
}
\label{table:depression}
\begin{tabular}{|l|cr|c} \hline
Name         &  \multicolumn{2}{c}{R$_{max}$}& $\Delta$([Fe/H])\\
             &  arcsec& pc                 &dex\\
\hline
FCC266       & 0.8  & 83         & 0.10 \\
FCC245       & 2.0  & 208        & 0.50 \\
NGC5898\_DW2 & 1.0  & 160        & 0.40 \\ 
FCC288       & 1.0  & 104        & 0.10 \\
FCC136       & 0.7  & 73         & 0.10 \\
\hline
\end{tabular}
\end{minipage}
\end{table}

Three of these objects are amongst those most dominated by an old
population (short formation time-scale; FCC 266 245 and 136) outside
of the depressed region their metallicity profile declines regularly,
and they are round objects.  The two others (FCC 288 and NGC5898\_DW2)
have flat metallicity profiles at large radii and contain an
intermediate age or young populations; the last one has H$_\beta$
emission, indicating residual ongoing star formation. They are flat
objects.

The metallicity depression can occur either in dE(di) objects or classical
dEs. It seems somewhat independent of the large
scale structure of the galaxy as well as the presence of a blue core
\citep{lt06} or nucleus \citep{cote06}.
Except for FCC245, whose peculiar profle is discussed in 
Appendix~\ref{appendix:individual}, the depressed region is not clearly
spatially resolved.

What interpretation to give to this phenomenon? As the seeing blurs
the data in a 100 pc region, we cannot test if it is due to the
characteristics of the nucleus or central cluster whose size is
of 10 pc or less. 

It is tempting to relate these depressions with the hypothesis that
the dE nuclei are large globular clusters \citep{zinnecker88}
or mergers of clusters which decayed to the centre \citep{lotz01}.

Other explanations may be proposed.  The mass of the nucleus is
roughly proportional to the total mass of the galaxy and continue to
low mass the relation between the masses of the super-massive
black-hole and of its host \citep{wehner06,ferrarese06,cote06}.  
The co-evolution that
the similarity of the two kinds of central-mass concentration would
imply, probably imposes that the nuclei started to form early from
low-metallicity gas and that its mean metallicity is at present
comparable to the mean metallicity of the galaxy.  Therefore, the
nuclear metallicity of the nucleus is expected to be lower than that
of the surrounding central region.

Another possibility
is that the central star formation is partly sustained by pristine
gas which lately cooled down. But the fact that depressions are observed
in some objects where we do not detect any recent star formation seems
to rule out this hypothesis.

\subsection{Origin of the dwarf elliptical galaxies} \label{sect:origin}

Although high-speed gravitational interactions in a dense cluster
environment can bring about the drastic morphological transformation
from discy, gas-rich spirals or dIrrs to rounder, gas-poor dEs \citep{m98,m05},
it remains to be seen whether the violent reshuffling of stellar
orbits and induced star formation indeed lead to dEs presenting
the characteristic metallicity gradients shown in this paper. 

Another, more plausible, explanation is that dE galaxies are
akin to dIrrs or BCDs but that the transformation is of a much more gentle
nature than the ``harassment'' process. Other gas removing processes
are ``starvation'' and ``strangulation'', suggested by a.o.
\cite{larson80,tt08,kamu08} and references therein. Once a gas-rich galaxy 
enters a cluster, the cluster potential imposes a Roche limit around this
galaxy, making some of the hot gas unbound. This gas is no longer
available for star formation. If the pressure of the intra-cluster
medium (ICM) is strong enough, it may help to strip away this outer
hot gas reservoir. Both processes are physically different but will
most likely work in unison in any cluster/group environment. The
outcome is in any case the same:~the supply of gas cooling from the
hot halo is cut off and the already accreted cold gas is transformed
into stars until none is left. As a consequence, star formation peters
out, leaving behind a gas-less dE with chemical and
kinematical properties very similar to those of a dIrr.

Ram-pressure stripping \citep{gg72,rh05,mmws06} provides another means
of removing gas from and consequently stopping star formation in
gas-rich dwarf galaxies accreting onto a galaxy cluster or
group. However, the efficiency of this mechanism, relative to
starvation/strangulation, depends critically on the density of the
intra-cluster or intra-group medium, a galaxy's velocity relative to
this medium, and the morphology of the infalling galaxy (round or
discy). Based on a suite of simulations of spherically symmetric dwarf
galaxies subjected to various ram-pressures, \citet{mori00} provide an
expression for the critical dark-matter core mass above which, for a
given ICM density and velocity, ram-pressure is not able to strip away
the gas. Evaluating this critical mass for different clusters, it is
clear that ram-pressure stripping of round dwarfs is a prominent
process in Virgo and Coma-like environments but not in the Fornax
cluster, the NGC5044 group or even less dense groups. Simulations
using dwarf disc galaxies, on the other hand, show that even
low-velocity, low-density ram-pressure is able to strip away the
lion's share of the gas in a time-scale of roughly 1~Gyr
\citep{marcolini03,hester06}. \citet{boselli08a, boselli08b} show that
the scaling relations of Virgo cluster dEs can be reproduced by
ram-ressure stripping of low-luminosity spiral galaxies or BCDs. These
authors could reject strangulation/starvation as the sole source of
these relations. No conclusion could be reached regarding harassment
because of the lack of enough N-body simulations of low-mass
galaxies. The correlation between the age of the stellar population,
estimated from absorption-line indices, and position within the Virgo
cluster, suggests that the gas is being removed quickly from the
infalling galaxies \citet{michielsen08b}.

It is clear that all these processes must act together in
group/cluster environments. Which process is the dominant agent in
removing gas from infalling dwarf galaxies depends on the
characteristics of the cluster/group.

\section{Conclusion} \label{conc}

In this paper we bring observations of 16 dEs that can be confronted 
with the different scenario of formation and evolution of galaxies:
\begin{enumerate}
\item
The star formation in the present dEs started in the early Universe.
The old population amounting for a dominant fraction of the mass is 
compatible with being co-eval of the oldest populations of massive
elliptical galaxies, or bulges.
\item
The star formation histories span a long time-scale. In general an
intermediate age population (1-5 Gyr) is present, and often a tail
of residual star formation extend to recent epochs.
The more recent episods of star formation are more concentrated than 
the old population.
\item
The youngest sub-populations have higher metallicities, and in general
the metallicity is decreasing from the centre outwards. The
gradients are typically 0.5 dex/\Reff. 
This seems to be a common properties of dEs, shared by NGC205 \citep{koleva09}.
These gradients are seen in the old (10 Gyr) population.
The steep metallicity gradients co-exist the age gradients
(older outside) due to the effect of concentrating the star 
formation, resulting is shallower color gradients.

\item
The flat and discy objects have almost constant metallicity. The
absence of gradient may either be due of their particular geometry
or may reflect a different origin like suggested by \citet{lt06}.
\item
We did not identify pecularities in the stellar populations of the 
central region of the 2 galaxies with kinematically decoupled cores.
But, in 5 galaxies we observe a depression of the metallicity
in the core, that may be the signature of a blue core diluted
by the seeing.
\end{enumerate}

What can we conclude about the origin of dEs and the forces driving
their evolution?

The prolonged star-formation history of dEs, 
compared with normal elliptical galaxies, is compatible with the idea of 
down-sizing 
(the shift of star formation to lower galaxy masses as cosmic time goes by).
The old population of dEs, co-eval to that of more massive galaxies,
may tell that the conditions to trigger the star formation
in intrinsically small galaxies were satified independently of
any strong environmental trigger. But alternatively, it may be compatible
with the tidal harassment being the origin of the dEs, if this process
can reproduce the other characteristics, and in particular the strong 
gradients. 

The metallicity gradients in the old population indicates that it
was not deeply mixed along its evolution. For instance,
the flat metallicity profiles of massive elliptical galaxies is
attributed to mergings: such an origin can certainly be excluded
for the dEs, but also, tidal harassments must not erase the gradients
if they are an important process to form dEs. The other environmental
effects that can quench the star formation by removing the gas
(ram-pressure stripping, starvation or strangulation
\citealp{gg72,rh05,mmws06,larson80,tt08,kamu08}) are more
gentle and more likely to preserve the gradients.

These metallicity gradients are produced in simulations 
\citep{pelupessy04,stinson07,valcke08}
as a result of the self-regulation of the star formation
by the stellar winds, but they generally settle on longer time
scales and were not expected in the old population.
At this point, more simulations would be required to try
to reproduce these metallicity gradients.

The geometry of these galaxies is also certainly a key factor
and is not well constrained by the present observations.
A better understanding of the core region and of the disc
galaxies requires to repeat stellar populations analyses
using IFU spectroscopy (as was made for one galaxies by
\citealp{cpsa07}).

The co-existence of various classes of dEs \citep{lt06,lt07}, the
discy dEs having flat metallicity profiles, 
may result from the transformation, by the same environmental effects,
a different progenitors, BCDs, dIrrs or small spiral galaxies.

>From this present work and from other recent observation emerge
a growingly consistent view of the formation of the diffuse dEs.
The gas depleted dEs were once star forming and gas rich. The gas was
exhausted by star formation, expelled by the winds and/or stripped by
ram-pressure against the intra-cluster medium or by tidal interaction
with other galaxies.  During their 'active' live, the star formation was
controlled by internal processes (feed-back) and by
environmental effects which can trigger a burst thanks to the
compression or finally finish it by cutting the gas supply.

\section*{Acknowledgments}

This paper is based on observations collected at the European Southern
Observatory, Paranal, Chile (programs 165.N-0115, 075.B-0179, and
076.B- 0196). MK acknowleges a PhD grant from the French Ambassy in Sofia.
SDR is a postdoctoral fellow with the National Science
Fund -- Flanders (FWO). This project was supported by a Tournesol
Scientific Exchange Programme between the Flemish Community and
France.

We are deeply indebted to Anna Pasquali, Victor P. Debattista and
Ignacio Ferreras who initiated this observational program. We thank
Pierre Ocvirk for providing publicly the \textsc{steckmap} package, and
Eric Emsellem for fruitful discussions. We also thank the anonymous referee
for his/her suggestions and comments.

\bibliographystyle{mn2e}
\bibliography{fors}   

\begin{thebibliography}{90}
\expandafter\ifx\csname natexlab\endcsname\relax\def\natexlab#1{#1}\fi

\bibitem[{{Alard}(2001)}]{alard01}
{Alard} C., 2001, \aap, 377, 389

\bibitem[{{Barazza} {et~al.}(2002){Barazza}, {Binggeli}, \& {Jerjen}}]{b02}
{Barazza} F.~D., {Binggeli} B., {Jerjen} H., 2002, \aap, 391, 823

\bibitem[{{Battaglia} {et~al.}(2006){Battaglia}, {Tolstoy}, {Helmi}, {Irwin},
  {Letarte}, {Jablonka}, {Hill}, {Venn}, {Shetrone}, {Arimoto}, {Primas},
  {Kaufer}, {Francois}, {Szeifert}, {Abel}, \& {Sadakane}}]{bthi06}
{Battaglia} G., {Tolstoy} E., {Helmi} A., {Irwin} M.~J., {Letarte} B.,
  {Jablonka} P., {Hill} V., {Venn} K.~A., {Shetrone} M.~D., {Arimoto} N.,
  {Primas} F., {Kaufer} A., {Francois} P., {Szeifert} T., {Abel} T., {Sadakane}
  K., 2006, \aap, 459, 423

\bibitem[{{Boselli} {et~al.}(2008{\natexlab{a}}){Boselli}, {Boissier},
  {Cortese}, \& {Gavazzi}}]{boselli08a}
{Boselli} A., {Boissier} S., {Cortese} L., {Gavazzi} G., 2008{\natexlab{a}},
  \apj, 674, 742

\bibitem[{{Boselli} {et~al.}(2008{\natexlab{b}}){Boselli}, {Boissier},
  {Cortese}, \& {Gavazzi}}]{boselli08b}
---, 2008{\natexlab{b}}, \aap, 489, 1015

\bibitem[{{Bouchard} {et~al.}(2007){Bouchard}, {Jerjen}, {Da Costa}, \&
  {Ott}}]{bouchard07}
{Bouchard} A., {Jerjen} H., {Da Costa} G.~S., {Ott} J., 2007, \aj, 133, 261

\bibitem[{{Buyle} {et~al.}(2005){Buyle}, {De Rijcke}, {Michielsen}, {Baes}, \&
  {Dejonghe}}]{Buyle05}
{Buyle} P., {De Rijcke} S., {Michielsen} D., {Baes} M., {Dejonghe} H., 2005,
  \mnras, 360, 853

\bibitem[{{Chilingarian} {et~al.}(2007){Chilingarian}, {Prugniel},
  {Sil'Chenko}, \& {Afanasiev}}]{cpsa07}
{Chilingarian} I.~V., {Prugniel} P., {Sil'Chenko} O.~K., {Afanasiev} V.~L.,
  2007, \mnras, 376, 1033

\bibitem[{{Cid Fernandes} {et~al.}(2005){Cid Fernandes}, {Mateus}, {Sodr{\'e}},
  {Stasi{\'n}ska}, \& {Gomes}}]{STARLIGHT}
{Cid Fernandes} R., {Mateus} A., {Sodr{\'e}} L., {Stasi{\'n}ska} G., {Gomes}
  J.~M., 2005, \mnras, 358, 363

\bibitem[{{Coelho} {et~al.}(2005){Coelho}, {Barbuy}, {Mel{\'e}ndez},
  {Schiavon}, \& {Castilho}}]{coelho05}
{Coelho} P., {Barbuy} B., {Mel{\'e}ndez} J., {Schiavon} R.~P., {Castilho}
  B.~V., 2005, \aap, 443, 735

\bibitem[{{Cole} {et~al.}(2007){Cole}, {Skillman}, {Tolstoy}, {Gallagher},
  {Aparicio}, {Dolphin}, {Gallart}, {Hidalgo}, {Saha}, {Stetson}, \&
  {Weisz}}]{cole07}
{Cole} A.~A., {Skillman} E.~D., {Tolstoy} E., {Gallagher} III J.~S., {Aparicio}
  A., {Dolphin} A.~E., {Gallart} C., {Hidalgo} S.~L., {Saha} A., {Stetson}
  P.~B., {Weisz} D.~R., 2007, \apjl, 659, L17

\bibitem[{{Conselice}(2006)}]{conselice06}
{Conselice} C.~J., 2006, \apj, 639, 120

\bibitem[{{C{\^o}t{\'e}} {et~al.}(2006){C{\^o}t{\'e}}, {Piatek}, {Ferrarese},
  {Jord{\'a}n}, {Merritt}, {Peng}, {Ha{\c s}egan}, {Blakeslee}, {Mei}, {West},
  {Milosavljevi{\'c}}, \& {Tonry}}]{cote06}
{C{\^o}t{\'e}} P., {Piatek} S., {Ferrarese} L., {Jord{\'a}n} A., {Merritt} D.,
  {Peng} E.~W., {Ha{\c s}egan} M., {Blakeslee} J.~P., {Mei} S., {West} M.~J.,
  {Milosavljevi{\'c}} M., {Tonry} J.~L., 2006, \apjs, 165, 57

\bibitem[{{Davidge}(2005)}]{davidge05}
{Davidge} T.~J., 2005, \aj, 130, 2087

\bibitem[{{De Rijcke} {et~al.}(2001){De Rijcke}, {Dejonghe}, {Zeilinger}, \&
  {Hau}}]{ddzh01}
{De Rijcke} S., {Dejonghe} H., {Zeilinger} W.~W., {Hau} G.~K.~T., 2001, \apjl,
  559, L21

\bibitem[{{De Rijcke} {et~al.}(2003{\natexlab{a}}){De Rijcke}, {Dejonghe},
  {Zeilinger}, \& {Hau}}]{d03}
---, 2003{\natexlab{a}}, \aap, 400, 119

\bibitem[{{De Rijcke} {et~al.}(2004){De Rijcke}, {Dejonghe}, {Zeilinger}, \&
  {Hau}}]{d04}
---, 2004, \aap, 426, 53

\bibitem[{{De Rijcke} {et~al.}(2005){De Rijcke}, {Michielsen}, {Dejonghe},
  {Zeilinger}, \& {Hau}}]{d05}
{De Rijcke} S., {Michielsen} D., {Dejonghe} H., {Zeilinger} W.~W., {Hau}
  G.~K.~T., 2005, \aap, 438, 491

\bibitem[{{De Rijcke} {et~al.}(2003{\natexlab{b}}){De Rijcke}, {Zeilinger},
  {Dejonghe}, \& {Hau}}]{dzdh03}
{De Rijcke} S., {Zeilinger} W.~W., {Dejonghe} H., {Hau} G.~K.~T.,
  2003{\natexlab{b}}, \mnras, 339, 225

\bibitem[{{Dolphin}(2002)}]{dolphin02}
{Dolphin} A.~E., 2002, \mnras, 332, 91

\bibitem[{{Dolphin} {et~al.}(2003){Dolphin}, {Saha}, {Skillman}, {Dohm-Palmer},
  {Tolstoy}, {Cole}, {Gallagher}, {Hoessel}, \& {Mateo}}]{dolphin03}
{Dolphin} A.~E., {Saha} A., {Skillman} E.~D., {Dohm-Palmer} R.~C., {Tolstoy}
  E., {Cole} A.~A., {Gallagher} J.~S., {Hoessel} J.~G., {Mateo} M., 2003, \aj,
  126, 187

\bibitem[{{Ferguson} \& {Binggeli}(1994)}]{fb94}
{Ferguson} H.~C., {Binggeli} B., 1994, \aapr, 6, 67

\bibitem[{{Ferrarese} {et~al.}(2006){Ferrarese}, {C{\^o}t{\'e}}, {Dalla
  Bont{\`a}}, {Peng}, {Merritt}, {Jord{\'a}n}, {Blakeslee}, {Ha{\c s}egan},
  {Mei}, {Piatek}, {Tonry}, \& {West}}]{ferrarese06}
{Ferrarese} L., {C{\^o}t{\'e}} P., {Dalla Bont{\`a}} E., {Peng} E.~W.,
  {Merritt} D., {Jord{\'a}n} A., {Blakeslee} J.~P., {Ha{\c s}egan} M., {Mei}
  S., {Piatek} S., {Tonry} J.~L., {West} M.~J., 2006, \apjl, 644, L21

\bibitem[{{Geha} {et~al.}(2003){Geha}, {Guhathakurta}, \& {van der
  Marel}}]{ggm03}
{Geha} M., {Guhathakurta} P., {van der Marel} R.~P., 2003, \aj, 126, 1794

\bibitem[{{Goto} {et~al.}(2003){Goto}, {Yamauchi}, {Fujita}, {Okamura},
  {Sekiguchi}, {Smail}, {Bernardi}, \& {Gomez}}]{gyf03}
{Goto} T., {Yamauchi} C., {Fujita} Y., {Okamura} S., {Sekiguchi} M., {Smail}
  I., {Bernardi} M., {Gomez} P.~L., 2003, \mnras, 346, 601

\bibitem[{{Gourgoulhon} {et~al.}(1992){Gourgoulhon}, {Chamaraux}, \&
  {Fouque}}]{gcf92}
{Gourgoulhon} E., {Chamaraux} P., {Fouque} P., 1992, \aap, 255, 69

\bibitem[{{Graham} \& {Guzm{\'a}n}(2003)}]{graham03}
{Graham} A.~W., {Guzm{\'a}n} R., 2003, \aj, 125, 2936

\bibitem[{{Graham} {et~al.}(2003){Graham}, {Jerjen}, \& {Guzm{\'a}n}}]{g03}
{Graham} A.~W., {Jerjen} H., {Guzm{\'a}n} R., 2003, \aj, 126, 1787

\bibitem[{{Grebel}(2000)}]{grebel00}
{Grebel} E.~K., 2000, in Bulletin of the American Astronomical Society,
  Vol.~32, pp. 698--+

\bibitem[{{Gunn} \& {Gott}(1972)}]{gg72}
{Gunn} J.~E., {Gott} J.~R.~I., 1972, \apj, 176, 1

\bibitem[{{Haines} {et~al.}(2007){Haines}, {Gargiulo}, {La Barbera},
  {Mercurio}, {Merluzzi}, \& {Busarello}}]{h07}
{Haines} C.~P., {Gargiulo} A., {La Barbera} F., {Mercurio} A., {Merluzzi} P.,
  {Busarello} G., 2007, \mnras, 381, 7

\bibitem[{{Heavens} {et~al.}(2000){Heavens}, {Jimenez}, \& {Lahav}}]{MOPED}
{Heavens} A.~F., {Jimenez} R., {Lahav} O., 2000, \mnras, 317, 965

\bibitem[{{Hester}(2006)}]{hester06}
{Hester} J.~A., 2006, \apj, 647, 910

\bibitem[{{Jerjen} {et~al.}(2000){Jerjen}, {Kalnajs}, \& {Binggeli}}]{j00}
{Jerjen} H., {Kalnajs} A., {Binggeli} B., 2000, \aap, 358, 845

\bibitem[{{Kawata} \& {Mulchaey}(2008)}]{kamu08}
{Kawata} D., {Mulchaey} J.~S., 2008, \apjl, 672, L103

\bibitem[{{Kent} {et~al.}(2007){Kent}, {Giovanelli}, {Haynes}, {Saintonge},
  {Stierwalt}, {Balonek}, {Brosch}, {Catinella}, {Koopmann}, {Momjian}, \&
  {Spekkens}}]{kent07}
{Kent} B.~R., {Giovanelli} R., {Haynes} M.~P., {Saintonge} A., {Stierwalt} S.,
  {Balonek} T., {Brosch} N., {Catinella} B., {Koopmann} R.~A., {Momjian} E.,
  {Spekkens} K., 2007, \apjl, 665, L15

\bibitem[{{Koleva}(2009)}]{koleva09}
{Koleva} M., 2009, PhD thesis, University of Lyon

\bibitem[{{Koleva} {et~al.}(2008{\natexlab{a}}){Koleva}, {Gupta}, {Prugniel},
  \& {Singh}}]{koleva08e}
{Koleva} M., {Gupta} R., {Prugniel} P., {Singh} H., 2008{\natexlab{a}}, in
  Astronomical Society of the Pacific Conference Series, Vol. 390, Pathways
  Through an Eclectic Universe, {Knapen} J.~H., {Mahoney} T.~J., {Vazdekis} A.,
  eds., pp. 302--+

\bibitem[{{Koleva} {et~al.}(2008{\natexlab{b}}){Koleva}, {Prugniel}, \& {De
  Rijcke}}]{koleva08b}
{Koleva} M., {Prugniel} P., {De Rijcke} S., 2008{\natexlab{b}}, Astronomische
  Nachrichten, 329, 968

\bibitem[{{Koleva} {et~al.}(2008{\natexlab{c}}){Koleva}, {Prugniel}, {Ocvirk},
  {Le Borgne}, \& {Soubiran}}]{koleva08a}
{Koleva} M., {Prugniel} P., {Ocvirk} P., {Le Borgne} D., {Soubiran} C.,
  2008{\natexlab{c}}, \mnras, 385, 1998

\bibitem[{{Kormendy}(1985)}]{kormendy85}
{Kormendy} J., 1985, \apj, 295, 73

\bibitem[{{Kormendy} {et~al.}(2008){Kormendy}, {Fisher}, {Cornell}, \&
  {Bender}}]{kormendy08}
{Kormendy} J., {Fisher} D.~B., {Cornell} M.~E., {Bender} R., 2008,
  arXiv:0810.1681

\bibitem[{{Kroupa} {et~al.}(1993){Kroupa}, {Tout}, \& {Gilmore}}]{kru93}
{Kroupa} P., {Tout} C.~A., {Gilmore} G., 1993, \mnras, 262, 545

\bibitem[{{Larson} {et~al.}(1980){Larson}, {Tinsley}, \& {Caldwell}}]{larson80}
{Larson} R.~B., {Tinsley} B.~M., {Caldwell} C.~N., 1980, \apj, 237, 692

\bibitem[{{Le Borgne} {et~al.}(2004){Le Borgne}, {Rocca-Volmerange},
  {Prugniel}, {Lan{\c c}on}, {Fioc}, \& {Soubiran}}]{PEGASEHR}
{Le Borgne} D., {Rocca-Volmerange} B., {Prugniel} P., {Lan{\c c}on} A., {Fioc}
  M., {Soubiran} C., 2004, \aap, 425, 881

\bibitem[{{Lee} {et~al.}(2006){Lee}, {Skillman}, \& {Venn}}]{lsv06}
{Lee} H., {Skillman} E.~D., {Venn} K.~A., 2006, \apj, 642, 813

\bibitem[{{Li} \& {Han}(2008)}]{li08}
{Li} Z., {Han} Z., 2008, \apj, 685, 225

\bibitem[{{Lisker} {et~al.}(2006){Lisker}, {Glatt}, {Westera}, \&
  {Grebel}}]{lt06}
{Lisker} T., {Glatt} K., {Westera} P., {Grebel} E.~K., 2006, \aj, 132, 2432

\bibitem[{{Lisker} {et~al.}(2007){Lisker}, {Grebel}, {Binggeli}, \&
  {Glatt}}]{lt07}
{Lisker} T., {Grebel} E.~K., {Binggeli} B., {Glatt} K., 2007, \apj, 660, 1186

\bibitem[{{Lotz} {et~al.}(2004){Lotz}, {Miller}, \& {Ferguson}}]{lotz04}
{Lotz} J.~M., {Miller} B.~W., {Ferguson} H.~C., 2004, \apj, 613, 262

\bibitem[{{Lotz} {et~al.}(2001){Lotz}, {Telford}, {Ferguson}, {Miller},
  {Stiavelli}, \& {Mack}}]{lotz01}
{Lotz} J.~M., {Telford} R., {Ferguson} H.~C., {Miller} B.~W., {Stiavelli} M.,
  {Mack} J., 2001, \apj, 552, 572

\bibitem[{{Mac Low} \& {Ferrara}(1999)}]{mafe99}
{Mac Low} M.-M., {Ferrara} A., 1999, \apj, 513, 142

\bibitem[{{Mapelli} {et~al.}(2007){Mapelli}, {Ripamonti}, {Tolstoy},
  {Sigurdsson}, {Irwin}, \& {Battaglia}}]{mapelli07}
{Mapelli} M., {Ripamonti} E., {Tolstoy} E., {Sigurdsson} S., {Irwin} M.~J.,
  {Battaglia} G., 2007, \mnras, 380, 1127

\bibitem[{{Marcolini} {et~al.}(2003){Marcolini}, {Brighenti}, \&
  {D'Ercole}}]{marcolini03}
{Marcolini} A., {Brighenti} F., {D'Ercole} A., 2003, \mnras, 345, 1329

\bibitem[{{Marcolini} {et~al.}(2008){Marcolini}, {D'Ercole}, {Battaglia}, \&
  {Gibson}}]{marcolini08}
{Marcolini} A., {D'Ercole} A., {Battaglia} G., {Gibson} B.~K., 2008, \mnras,
  386, 2173

\bibitem[{{Mart{\'{\i}}nez-Delgado} {et~al.}(1999){Mart{\'{\i}}nez-Delgado},
  {Aparicio}, \& {Gallart}}]{martinez-delgado99}
{Mart{\'{\i}}nez-Delgado} D., {Aparicio} A., {Gallart} C., 1999, \aj, 118, 2229

\bibitem[{{Mastropietro} {et~al.}(2005){Mastropietro}, {Moore}, {Mayer},
  {Debattista}, {Piffaretti}, \& {Stadel}}]{m05}
{Mastropietro} C., {Moore} B., {Mayer} L., {Debattista} V.~P., {Piffaretti} R.,
  {Stadel} J., 2005, \mnras, 364, 607

\bibitem[{{Mayer} {et~al.}(2001){Mayer}, {Governato}, {Colpi}, {Moore},
  {Quinn}, {Wadsley}, {Stadel}, \& {Lake}}]{m01}
{Mayer} L., {Governato} F., {Colpi} M., {Moore} B., {Quinn} T., {Wadsley} J.,
  {Stadel} J., {Lake} G., 2001, \apjl, 547, L123

\bibitem[{{Mayer} {et~al.}(2006){Mayer}, {Mastropietro}, {Wadsley}, {Stadel},
  \& {Moore}}]{mmws06}
{Mayer} L., {Mastropietro} C., {Wadsley} J., {Stadel} J., {Moore} B., 2006,
  \mnras, 369, 1021

\bibitem[{{Michielsen} {et~al.}(2008){Michielsen}, {Boselli}, {Conselice},
  {Toloba}, {Whiley}, {Arag{\'o}n-Salamanca}, {Balcells}, {Cardiel}, {Cenarro},
  {Gorgas}, {Peletier}, \& {Vazdekis}}]{michielsen08b}
{Michielsen} D., {Boselli} A., {Conselice} C.~J., {Toloba} E., {Whiley} I.~M.,
  {Arag{\'o}n-Salamanca} A., {Balcells} M., {Cardiel} N., {Cenarro} A.~J.,
  {Gorgas} J., {Peletier} R.~F., {Vazdekis} A., 2008, \mnras, 385, 1374

\bibitem[{{Michielsen} {et~al.}(2004){Michielsen}, {De Rijcke}, {Zeilinger},
  {Prugniel}, {Dejonghe}, \& {Roberts}}]{mdzpdr04}
{Michielsen} D., {De Rijcke} S., {Zeilinger} W.~W., {Prugniel} P., {Dejonghe}
  H., {Roberts} S., 2004, \mnras, 353, 1293

\bibitem[{{Michielsen} {et~al.}(2007){Michielsen}, {Koleva}, {Prugniel},
  {Zeilinger}, {De Rijcke}, {Dejonghe}, {Pasquali}, {Ferreras}, \&
  {Debattista}}]{michielsen08}
{Michielsen} D., {Koleva} M., {Prugniel} P., {Zeilinger} W.~W., {De Rijcke} S.,
  {Dejonghe} H., {Pasquali} A., {Ferreras} I., {Debattista} V.~P., 2007, \apjl,
  670, L101

\bibitem[{{Moore} {et~al.}(1998){Moore}, {Lake}, \& {Katz}}]{m98}
{Moore} B., {Lake} G., {Katz} N., 1998, \apj, 495, 139

\bibitem[{{Mori} \& {Burkert}(2000)}]{mori00}
{Mori} M., {Burkert} A., 2000, \apj, 538, 559

\bibitem[{{Moultaka}(2005)}]{moultaka05}
{Moultaka} J., 2005, \aap, 430, 95

\bibitem[{{Nieto} \& {Prugniel}(1987)}]{np87}
{Nieto} J.-L., {Prugniel} P., 1987, \aap, 186, 30

\bibitem[{{Ocvirk} {et~al.}(2006){Ocvirk}, {Pichon}, {Lan{\c c}on}, \&
  {Thi{\'e}baut}}]{ocv06}
{Ocvirk} P., {Pichon} C., {Lan{\c c}on} A., {Thi{\'e}baut} E., 2006, \mnras,
  365, 74

\bibitem[{{Pasquali} {et~al.}(2005){Pasquali}, {Larsen}, {Ferreras}, {Gnedin},
  {Malhotra}, {Rhoads}, {Pirzkal}, \& {Walsh}}]{pasquali05}
{Pasquali} A., {Larsen} S., {Ferreras} I., {Gnedin} O.~Y., {Malhotra} S.,
  {Rhoads} J.~E., {Pirzkal} N., {Walsh} J.~R., 2005, \aj, 129, 148

\bibitem[{{Paturel} {et~al.}(2003){Paturel}, {Petit}, {Prugniel}, {Theureau},
  {Rousseau}, {Brouty}, {Dubois}, \& {Cambr{\'e}sy}}]{HYPERLEDA}
{Paturel} G., {Petit} C., {Prugniel} P., {Theureau} G., {Rousseau} J., {Brouty}
  M., {Dubois} P., {Cambr{\'e}sy} L., 2003, \aap, 412, 45

\bibitem[{{Pelupessy} {et~al.}(2004){Pelupessy}, {van der Werf}, \&
  {Icke}}]{pelupessy04}
{Pelupessy} F.~I., {van der Werf} P.~P., {Icke} V., 2004, \aap, 422, 55

\bibitem[{{Prugniel} {et~al.}(1999){Prugniel}, {Golev}, \&
  {Maubon}}]{prugniel99}
{Prugniel} P., {Golev} V., {Maubon} G., 1999, \aap, 346, L25

\bibitem[{{Prugniel} {et~al.}(2007{\natexlab{a}}){Prugniel}, {Koleva},
  {Ocvirk}, {Le Borgne}, \& {Soubiran}}]{prugniel07}
{Prugniel} P., {Koleva} M., {Ocvirk} P., {Le Borgne} D., {Soubiran} C.,
  2007{\natexlab{a}}, in IAU Symposium, Vol. 241, IAU Symposium, {Vazdekis} A.,
  {Peletier} R.~F., eds., pp. 68--72

\bibitem[{{Prugniel} \& {Simien}(2003)}]{ps023}
{Prugniel} P., {Simien} F., 2003, \apss, 284, 603

\bibitem[{{Prugniel} \& {Soubiran}(2001)}]{PS01}
{Prugniel} P., {Soubiran} C., 2001, \aap, 369, 1048

\bibitem[{{Prugniel} {et~al.}(2007{\natexlab{b}}){Prugniel}, {Soubiran},
  {Koleva}, \& {Le Borgne}}]{ELODIE31}
{Prugniel} P., {Soubiran} C., {Koleva} M., {Le Borgne} D., 2007{\natexlab{b}},
  arXiv:astro-ph/0703658

\bibitem[{{Rieschick} \& {Hensler}(2003)}]{riehe03}
{Rieschick} A., {Hensler} G., 2003, \apss, 284, 861

\bibitem[{{Roediger} \& {Hensler}(2005)}]{rh05}
{Roediger} E., {Hensler} G., 2005, \aap, 433, 875

\bibitem[{{S{\'a}nchez-Bl{\'a}zquez} {et~al.}(2006){S{\'a}nchez-Bl{\'a}zquez},
  {Peletier}, {Jim{\'e}nez-Vicente}, {Cardiel}, {Cenarro},
  {Falc{\'o}n-Barroso}, {Gorgas}, {Selam}, \& {Vazdekis}}]{so6}
{S{\'a}nchez-Bl{\'a}zquez} P., {Peletier} R.~F., {Jim{\'e}nez-Vicente} J.,
  {Cardiel} N., {Cenarro} A.~J., {Falc{\'o}n-Barroso} J., {Gorgas} J., {Selam}
  S., {Vazdekis} A., 2006, \mnras, 371, 703

\bibitem[{{Stinson} {et~al.}(2007){Stinson}, {Dalcanton}, {Quinn}, {Kaufmann},
  \& {Wadsley}}]{stinson07}
{Stinson} G.~S., {Dalcanton} J.~J., {Quinn} T., {Kaufmann} T., {Wadsley} J.,
  2007, \apj, 667, 170

\bibitem[{{Thomas} {et~al.}(2003{\natexlab{a}}){Thomas}, {Bender}, {Hopp},
  {Maraston}, \& {Greggio}}]{thomas03b}
{Thomas} D., {Bender} R., {Hopp} U., {Maraston} C., {Greggio} L.,
  2003{\natexlab{a}}, \apss, 284, 599

\bibitem[{{Thomas} {et~al.}(2003{\natexlab{b}}){Thomas}, {Maraston}, \&
  {Bender}}]{thomas03}
{Thomas} D., {Maraston} C., {Bender} R., 2003{\natexlab{b}}, \mnras, 339, 897

\bibitem[{{Tojeiro} {et~al.}(2007){Tojeiro}, {Heavens}, {Jimenez}, \&
  {Panter}}]{VESTA}
{Tojeiro} R., {Heavens} A.~F., {Jimenez} R., {Panter} B., 2007, \mnras, 381,
  1252

\bibitem[{{Tully} \& {Trentham}(2008)}]{tt08}
{Tully} R.~B., {Trentham} N., 2008, \aj, 135, 1488

\bibitem[{{Valcke} {et~al.}(2008){Valcke}, {de Rijcke}, \&
  {Dejonghe}}]{valcke08}
{Valcke} S., {de Rijcke} S., {Dejonghe} H., 2008, \mnras, 389, 1111

\bibitem[{{van den Bosch} {et~al.}(2008){van den Bosch}, {Aquino}, {Yang},
  {Mo}, {Pasquali}, {McIntosh}, {Weinmann}, \& {Kang}}]{vandenbosch08}
{van den Bosch} F.~C., {Aquino} D., {Yang} X., {Mo} H.~J., {Pasquali} A.,
  {McIntosh} D.~H., {Weinmann} S.~M., {Kang} X., 2008, \mnras, 387, 79

\bibitem[{{van Zee} {et~al.}(2004){van Zee}, {Skillman}, \& {Haynes}}]{vz04}
{van Zee} L., {Skillman} E.~D., {Haynes} M.~P., 2004, \aj, 128, 121

\bibitem[{{Wehner} \& {Harris}(2006)}]{wehner06}
{Wehner} E.~H., {Harris} W.~E., 2006, \apjl, 644, L17

\bibitem[{{Wheeler} {et~al.}(1989){Wheeler}, {Sneden}, \& {Truran}}]{wheeler89}
{Wheeler} J.~C., {Sneden} C., {Truran} Jr. J.~W., 1989, \araa, 27, 279

\bibitem[{{Young} \& {Lo}(1997)}]{yl97}
{Young} L.~M., {Lo} K.~Y., 1997, \apj, 476, 127

\bibitem[{{Zinnecker} {et~al.}(1988){Zinnecker}, {Keable}, {Dunlop}, {Cannon},
  \& {Griffiths}}]{zinnecker88}
{Zinnecker} H., {Keable} C.~J., {Dunlop} J.~S., {Cannon} R.~D., {Griffiths}
  W.~K., 1988, in IAU Symposium, Vol. 126, The Harlow-Shapley Symposium on
  Globular Cluster Systems in Galaxies, {Grindlay} J.~E., {Philip} A.~G.~D.,
  eds., pp. 603--+

\end{thebibliography}

\appendix

\section[]{Galaxy by galaxy description} \label{appendix:individual}
This Appendix gathers comments on individual objects written down during
the spectroscopic analysis. We briefly discuss the
of the SFH in 1~arcsec and 1R$_e$, and the SSP-equivalent parameters
of the 'corrected' from the light on the line-of-sight inner regions
in each object. Except
when explicitly noted, we chose the SFH on the basis of the 
goodness and stability of the fit.
Usually there was no particular spectral features which drove our
choice of the SFH. 
It was more matter of general smaller residuals (i. e. smaller \csq).

\medskip\noindent{\bf FCC043}.
The inner 1~arcsec of FCC043 is well described with one coeval population at
intermediate age and metallicity. In the outer part (1\Reff{} extraction)
we find a metal poor old component (30per cent in light). We were unable
to subtract the contamination of the light on the line-of-sight from
the 1~arcsec extraction.

The spectrum is contaminated by a background object about 20~arcsec
East of the centre: The emission line at 4488~{\AA} (in the galaxy
rest-frame) seen between 17 and 24~arcsec is probably 
[O{\sc iii}]~$\lambda 3727${\AA} redshifted at z=0.21. The residuals show
H$_\delta$ in absorption and H$_\gamma$ in absorption plus
emission. We excluded the contaminated spectra from the analysis.

 \medskip\noindent{\bf FCC046}
This galaxy has emission in H$_{\beta}$ and [O{\sc iii}].
It is nucleated and is detected in HI. 
We could distinguish 4 populations in both the 1~arcsec and 1\Reff{} extractions.
The light fraction of the oldest population is
20-22per cent (for the 1~arcsec and 1\Reff{} extractions respectively).
Its star formation history reveals a strong bursts at 12~Gyr (much metal poor),
small residual star-formation during the middle ages and a new
present strong star formation with solar metallicity, more prominent 
in the centre. Another indication of the mix of young and old population
is the HeI 4388{\AA} line, due to hot stars, which is better fitted in the case of 
multiple SSPs.
The correction from the light on the line-of-sight reveals more metal-poor
population.

\medskip\noindent {\bf FCC136}
This flat, S0 galaxy has a small amount of intermediate
stars in the centre. The SSP-equivalent ages and metallicities
of the corrected spectrum are smaller than in the 1~arcsec extraction.
In the outer part we could not detect anything but old population. 

\medskip\noindent {\bf FCC150}
The inner 1~arcsec of this galaxy displays old and intermediate component.
The corrected inner part is slightly more metal-poor and younger than
the uncorrected 1~arcsec parameters. Curiously we were able
to detect young component in the outer region, but not in the inner.

\medskip\noindent {\bf FCC204}
 This is a galaxy with a spiral structure. The SFH in the both 1\Reff{}
and 1~arcsec extractions are consistent with two subpopulations
of old and intermediate ages.

\medskip\noindent {\bf FCC207}
The inner part of this galaxy displays constant SFH till recent
epochs. The presence of a young population is marked by
Balmer and [O{\sc iii}] emissions. In the 1\Reff{}
region we failed to find a young component. 

\begin{figure}
\includegraphics[width=0.98\columnwidth]{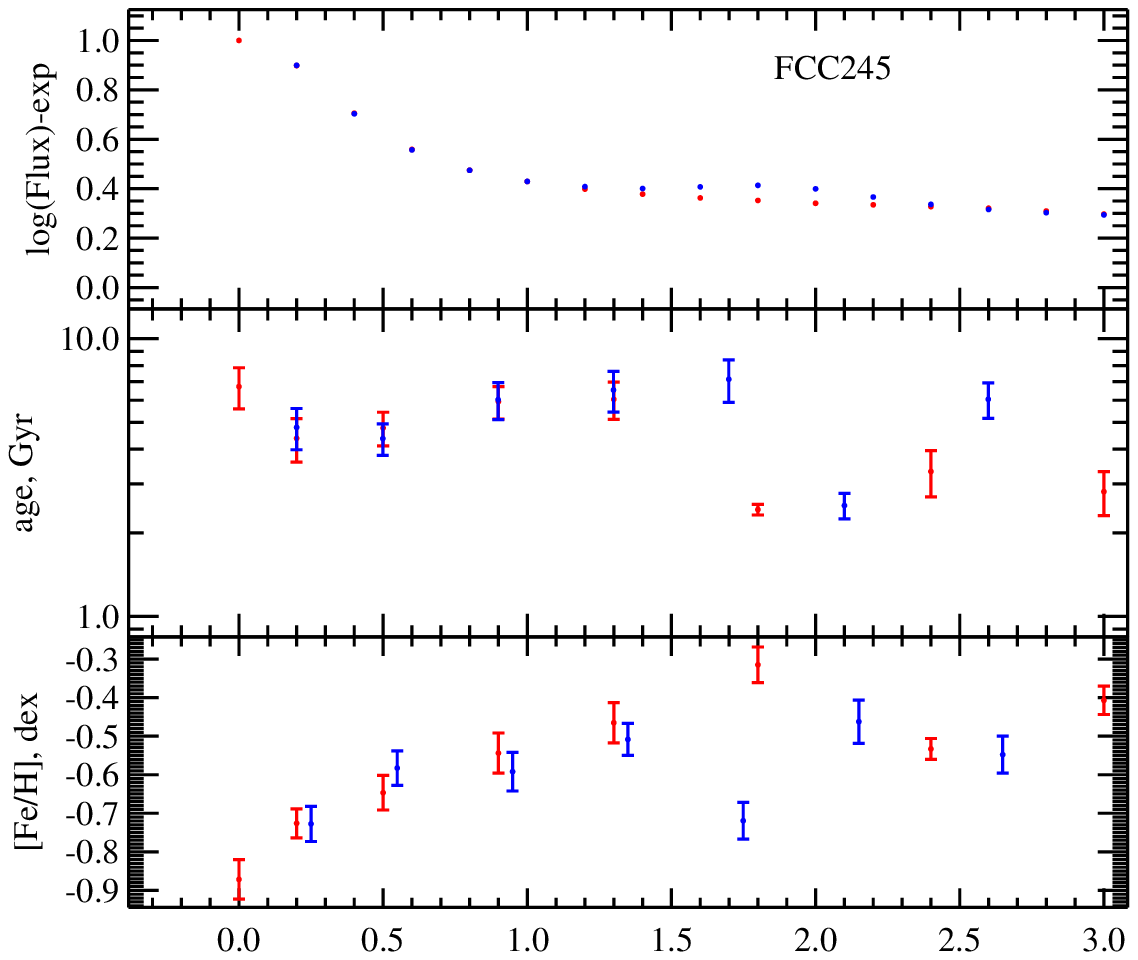}
\caption{Central SSP-equivalent profiles of FCC245.
The top panel is the excess over an exponential light profile fitted
using the long-slit spectra in the region between 2 and 5~arcsec.
The symbols and binning are identical to Fig.~\ref{fig:depression}.
}
\label{fig:fcc245_center}
\end{figure}

\medskip\noindent {\bf FCC245}
This galaxy is best fitted with a two-population (intermediate
and old) scenario. Its corrected core shows a decline in the SSP-equivalent 
metallicity and age.
The SSP profiles of FCC245 are peculiar. It is the galaxy with 
the best evidence for a central metallicity
depression.
The metallicity in the centre is about -0.8~dex, it is peaking at -0.5~dex
between $r=1.4$ and 4~arcsec, and 
decreases after to -0.8 again (Fig.~\ref{fig:fcc245_center}).
The SFH indicates the presence of an old population of metallicity 
-1~dex with an intermediate one overimposed.

We extracted the spectra in $r = $1.3 to 4~arcsec and studied the SFH. 
In order to ease the comparison with the 
other extractions we fixed the old population to 12 Gyr, [Fe/H]=-1~dex.
We find almost the same population mix in the central and 1\Reff{} 
extractions, with an intermediate age population
aged about 1.5 Gyr and a super-solar metallicity [Fe/H]=+0.4~dex amounting 
25per cent of the light. In the middle extraction the intermediate population 
is also the same, but has a fraction of 35 per cent of the light.
This SFH accounts reasonably for the observed SSP profiles.

It looks like if this dE0 has a ring with an enhanced 
intermediate population. But the photometry does not provide 
any hint for such a feature.

\medskip\noindent {\bf FCC266}
This is a nucleated dE. We detect a small fraction of young 
population in 1~arcsec
and 1\Reff, but it is going to the age low limit. We adopted a
 2-bursts fit. When subtracting the line-of-sight contamination of the
light we was left with only noise and we could not fit any SSP to it.

The North side of the galaxy spectrum is contaminated by a background
barred spiral galaxy located at red-shift 0.23, having an
SSP-equivalent age of 2.5 Gyr and solar metallicity. We did not
attempt to decontaminate the spectra and simply excluded the region
beyond 5~arcsec on that side.

\medskip\noindent {\bf FCC288}
A spiral structure is found in FCC288. In the outer part we detect a
 young population which is not seen in the inner 1~arcsec
(like in the case of FCC150).
In the both extractions the Mg$_b$ is over-fitted, which suggests slight
$\alpha$-under-abundance in comparing to our models (with solar
neighborhood abundance, \citealt{wheeler89}). 

\medskip\noindent {\bf DW1}
In the inner arcsecond the two-components population (old plus intermediate age)
give the best and more stable fit (in \csq{} sense). 
For the 1\Reff{} the situation is more complex
since if we adopt the three population fit the young component
goes to the high metallicity limit. On the other side, the fit is
worse when only two populations are considered. So, even if with 
an uncomfortably high metallicity we keep the three bursts model.
The corrected inner part have lower SSP-equivalent metallicity and age
in comparision with the non-corrected one burst fit to 1~arcsec.

\medskip\noindent {\bf DW2}
This galaxy has a similar SFH as FCC046. Actively star forming at
old ages and at present and more quiet at intermediate ages. The
SSP-equivalent fit is considerably worse with strong residuals at
G4300, Fe and Mg lines. A SSP obviously does not match the observation.
In the corrected 1~arcsec we find an young population with low 
metallicity.

\medskip\noindent {\bf FS29}
The best SFH for this galaxy seems to be represented by an intermediate
and old burst. In central 1~arcsec the star formation seems to be more
constant, while in the outer part it is rapidly diminishing.

 \medskip\noindent{\bf FS75}
The outer 1\Reff shows a small component but the metallicity is not constrained
(too big error bars) and we adopted the two bursts fit. Indeed it also
has strong  Balmer and oxygen emission lines.
It is one of the rare cases, where the corrected 1~arcsec is more
metal-rich than the uncorrected.

\medskip\noindent {\bf FS131}
 This is a peanut shaped galaxy which is also marked as nucleated.
We could not resolve more than one population in the central 1~arcsec,
even when we tried to correct for the outside light contamination.
In the outer regions however we see some fast decreasing SFH,
composed from an old and intermediate population.

\medskip\noindent {\bf FS373}
This galaxy was resolved in 3 populations. While in the
centre the intermediate age burst is dominating the mass, it has
the smallest contribution in the outskirts.
Both extractions show emission in [O{\sc ii}]. The S/N was not
sufficient to obtain SSP parameters for the corrected 1~arcsec.

\label{lastpage}

\end{document}